\documentclass[twocolumn,]{aastex62}
\usepackage{graphicx,color}        
\usepackage{color}            
\usepackage{amsmath}            
\usepackage{psfig}
\usepackage{subfloat,subfigure}
\usepackage{natbib}
\usepackage{hyperref} 
\hypersetup{linkcolor=red,citecolor=blue,filecolor=cyan,urlcolor=magenta}

\newcommand{\zerosix}{SDSS J064301.86+593430.9~}
\newcommand{\zerosixgc}{SDSS J064655.6+411620.5~}
\newcommand{\sixfiftytwo}{SDSS J065252.76+410506.0~}
\newcommand{\ten}{SDSS J102411.84+415146.7~}

\newcommand{\nineteengc}{SDSS J193712.01+502455.5~}
\newcommand{\sdssfiftythree}{SDSS J195344.22+422249.9~}

\accepted{August 11, 2022}

\submitjournal{ApJ}

\shorttitle{Li distribution for VMP stars}
\shortauthors{A. Bandyopadhyay et al.}

\begin{document}

\title{Li distribution, kinematics and detailed abundance analysis among very metal-poor stars in the Galactic halo from the HESP-GOMPA survey}

\correspondingauthor{Avrajit Bandyopadhyay}
\email{avrajit.india@gmail.com, avrajit@aries.res.in}

\author[0000-0002-8304-5444]{Avrajit Bandyopadhyay}
\affiliation{Aryabhatta Research Institute of Observational Sciences, Nainital 263001, India}

\author{Thirupathi Sivarani}
\affiliation{Indian Institute of Astrophysics, Bangalore, India}

\author[0000-0003-4573-6233]{Timothy C. Beers}
\affiliation{Department of Physics and Astronomy and JINA Center for the Evolution of the Elements, University of Notre Dame, Notre Dame, IN 46556 USA}

\author{A. Susmitha}
\affiliation{Indian Institute of Astrophysics, Bangalore, India}

\author[0000-0002-4638-1035]{Prasanta K Nayak}
\affiliation{Tata Institute of Fundamental Research, Colaba, Mumbai, 400005, India}

\author[0000-0002-4331-1867]{Jeewan C Pandey}
\affiliation{Aryabhatta Research Institute of Observational Sciences, Nainital 263001, India}

\begin{abstract}

We present a study on the detailed elemental abundances of newly identified bright very metal-poor stars with the detection of lithium, initially observed as part of the SDSS/MARVELS pre-survey. These stars were selected for high-resolution spectroscopic follow-up as part of the HESP-GOMPA survey. In this work, we discuss the Li abundances detected for several stars in the survey, which include main-sequence stars, subgiants, and red giants. Different classes of stars are found to exhibit very similar distributions of Li, which points towards a common origin. We derive a scaling relation for the depletion of Li as a function of temperature for giants and main-sequence stars; the majority of the samples from the literature were found to fall within 1$\sigma$  (0.19 and 0.12 dex/K for giants and dwarfs respectively) of this relationship. We also report the existence of a slope of the Li abundances as a function of distances from the Galactic plane, indicating mixed stellar populations. Most Li-rich stars are found to be in or close to the galactic plane. Along with Li, we have derived detailed abundances for C, odd-Z, $\alpha$-, Fe-peak and neutron-capture elements for each star.  We have also used astrometric parameters from Gaia-EDR3 to complement our study, and derived kinematics to differentiate between the motions of the stars; those formed in situ and accreted. The stellar population of the Spite plateau, including additional stars from the literature, is found to have significant contributions from stars formed in situ and through accretion. The orbits for the program stars have also been derived and studied for a period of 5 Gyr backwards in time.
 
\end{abstract}


\section{Introduction} \label{sec:intro}

The discovery of large numbers of very metal-poor(VMP; [Fe/H\ $< -2.0$) stars has provided great opportunities to study the pristine conditions that existed in the early Universe when these old stellar objects were formed \citep{beers2005,firststars6,frebelandnorris,frebelrev18}. Among the many studies that could be conducted with these stars, the detection and measurement of lithium are of particular importance. Lithium is the only element in the periodic table, apart from H and He, that owes its origin (at least in part) to Big-Bang nucleosynthesis. All other elements can be produced in stellar interiors or other exotic stellar events.

Lithium is also a very fragile element and is easily destroyed when exposed to higher temperatures, which can be inferred from the observed depletion of stellar Li content as a star ascends the giant branch and the stellar atmosphere is mixed with Li-depleted matter from the stellar interior, due to the convective channels that are opened during this phase. This so-called evolutionary mixing largely depletes lithium, lowering its observed absolute abundance, $A$(Li).

The pioneering study of \citet{spitenspite} reported the abundance of Li for a sample of the unevolved, older population of stars in the halo and disc of the Milky Way. A constant abundance of Li, $A$(Li) = 2.2, was obtained, and subsequently referred to as the "Spite Li plateau".  Over time, it came to be recognized that this level was substantially lower than the cosmological predictions of $A$(Li) = 2.7, based on the baryon density determined by the CMB measurements of the WMAP satellite  \citep{spergel2003,coc2004}. This discrepancy demonstrates the existence of physical processes that have resulted in the depletion of Li in metal-poor main-sequence turnoff (MSTO) stars. Since then, there have been many studies and efforts to understand the Li plateau and solve the Li problem (e.g., \citealt{Pinsonneault, ryan2002, korn2006, piau2006, firststars7}, among many others). A small, but statistically significant slope of the Li plateau was discovered by \citet{ryan1999} as more stars with Li detection were studied. The decreasing trend of Li abundances with a decrease in metallicity was confirmed by \cite{bonifacio2007} and \citet{sbordone2010}.  Extremely metal-poor(EMP; [Fe/H] $<-3.0$) were also found to have Li abundances lower than the Spite plateau (e.g., \citealt{bonifacio2015}, and references therein), causing the "breakdown'' \citep{aoki2009} or "meltdown'' of the Spite plateau \citep{sbordone2010}. 

Apart from Li, the abundances of other important elements among VMP and EMP stars are of considerable interest for constraining the pollution of their natal gas clouds by previous stellar generations. They also provide valuable constraints for improvements in models of stellar nucleosynthesis. Additional discoveries of, in particular, bright stars with [Fe/H] $<-2.0$, with or without chemical anomalies, are crucial for a better understanding of the nature of nucleosynthetic events in the early Universe.

In this paper, we report Li abundances for 12 metal-poor stars (including 10 VMP stars and 1 EMP star), 9 of which are studied for the first time. We have included three stars with measured Li reported earlier, and use them for investigations of their kinematics. The kinematics of these three stars were not reported previously and are included here for the sake of completeness of this chemo-dynamical study of Li abundances in stars from HESP-GOMPA survey. In Section 2, we describe the target selection and details of the high-resolution spectroscopic observations. Derivations of stellar parameters and the measurement of Li abundances are described in Section 3. Implications of these measurements, possible correlations with atmospheric parameters and other abundances, and the kinematics of our sample, supplemented by literature studies, are described in Section 4. Section 5 presents a brief summary and conclusions.

\section{Observations, target selection, and analysis} 

High-resolution ($R \sim 30,000$ \& $60,000$) spectroscopic observations of our program stars were carried out as a part of the HESP-GOMPA (Hanle Echelle SPectrograph -- Galactic survey Of metal-poor stArs) survey, using the HESP \citep{sriram2018} on the 2-m Himalayan Chandra Telescope (HCT) at the Indian Astronomical Observatory (IAO). The targets were selected from the spectroscopic pre-survey for MARVELS \citep{ge2015}, which was carried out as a part of SDSS-III \citep{eisenstein}. This offers the chance to identify bright metal-poor halo stars which could be studied at high spectral resolution using moderate-aperture telescopes. 
We have used synthetic spectral fitting of the pre-survey data to identify the most metal-poor stars. Furthermore, the metal-poor candidates with weak CH $G$-bands were given preference for the high-resolution follow-up observations to remove the carbon-rich stellar populations. We have obtained high-resolution data for 60 metal-poor stars, out of which Li could be measured for the 12 program stars listed in Table 1.  In this paper, there are 9 new stars with measured Li abundances, however, the abundance table has not been included for \sdssfiftythree. This object is a CEMP-no star, which will be discussed in an upcoming paper on CEMP stars (Bandyopadhyay et al., in prep.). Abundances for the remaining 8 stars are discussed below. Complete details for the others, including all of the observed stars in the HESP-GOMPA survey, will be discussed in a separate paper (Bandyopadhyay et al., in prep). The stars were observed at a spectral resolving power of $R \sim$ 30,000, spanning a wavelength range of 380 nm to 1000 nm. The coordinates and observation details, including duration of observation, signal-to-noise ratios, $V$ magnitudes, and radial velocities for the program stars are listed in Table 1.

\begin{table*}
\tabcolsep7.0pt $ $
\begin{center}
\caption{Observational Details for the Program Stars}
\begin{tabular}{ccccccccr}
\hline\hline
Star name &Object &RA &DEC &Exp & $SNR$ &$V$ mag & \r RV\\
 & &J(2000) &J(2000) & (sec) &  &  &(km/s) \\
\hline
SDSS J002400.64+320311.4 &SDSS J0024+3203 &00 24 00.64  &32 03 11.40 &7200 &70.3 &11.58 &  $-$434.0\\%
SDSS J031522.0+212324.6 &SDSS J0315+2123 &03 15 22.00  &21 23 24.60 &7200 &55.4 &11.35 & $-$49.0 \\%
SDSS J064301.86+593430.9 &SDSS J0643+5934 &06 43 01.86  &59 34 30.90 &8100 &71.6 &11.44 & 52.2\\%
SDSS J064655.60+411620.5 &SDSS J0646+4116 &06 46 55.60  &41 16 20.50 &9600 &43.1 &11.14 & $-$285.0\\
SDSS J065252.76+410506.0 &SDSS J0652+4105 &06 52 52.76  &41 05 06.00 &8100 &68.0 &11.36 &98.5\\%
SDSS J102411.84+415146.7 &SDSS J1024+4151 &10 24 11.84  &41 51 46.70 &7200 &49.5 &11.83 &194.0\\%
SDSS J114658.70+234357.2 &SDSS J1146+2343 &11 46 58.70  &23 43 57.20 &8100 &49.1 &11.06 & $-$9.5 \\%
SDSS J134144.60+474128.9 &SDSS J1341+4741 &13 41 44.60  &47 41 28.90 &7200 &47.0 &12.38 & $-$190.5\\
SDSS J172548.56+420241.9 &SDSS J1725+4202 &17 25 48.56  &42 02 41.90  &8100 &53.0 &11.66 & $-$266.5\\%
SDSS J193344.73+452410.9 &SDSS J1933+4524 &19 33 44.73  &45 24 10.90 &8100 &65.5 &11.48 &157.0\\
SDSS J193712.01+502455.5 &SDSS J1937+5024 &19 37 12.01  &50 24 55.50 &7200 &130.0 &10.44 & $-$184.0\\
SDSS J195344.22+422249.9 &SDSS J1953+4222 &19 53 44.22  &42 22 49.90 &7200 &245.0 &9.23 & $-$308.1\\

\hline
\end{tabular}
 \end{center}
\end{table*}

Data reduction was carried out using the IRAF echelle package, along with the publicly available data reduction pipeline for HESP\footnote{https://www.iiap.res.in/hesp/}, developed by Arun Surya. A cross-correlation analysis with a synthetic template spectrum was carried out to obtain the radial velocity (RV) for each star. The calculated RVs are listed in Table 1.

We have employed one-dimensional LTE stellar atmospheric models (ATLAS9; \citealt{castellikurucz}) and the spectral synthesis code TURBOSPECTRUM \citep{alvarezplez1998} for determining the abundances of the individual elements present in each spectrum. We have considered the equivalent widths of the absorption lines present in the spectra that are less than 120 m{\AA}, as they are on the linear part of the curve of growth. Version 12 of the TURBOSPECTRUM code for spectrum synthesis and abundance estimates was used for the analysis. The Kurucz database \footnote{http://kurucz.harvard.edu/linelists.html} was used for the compilation of the linelist. We have adopted the hyperfine splitting provided by \cite{mcwilliam1998}, along with Solar isotopic ratios.

The stellar atmospheric parameters of the program stars were derived iteratively. The first estimates for effective temperature were made using photometric colours, $V-K$. Gaia and SED \citep{bayo2008vosa} were also used to derive the values of $T_{\rm eff}$ and $\log g$. A grid for stellar models was prepared for a wide range of $T_{\rm eff}$, $\log g$, and [Fe/H]. The abundances of the clean Fe I and Fe II lines were measured for each spectrum by the method of equivalent-width analysis. The best fit was determined so that Fe I abundances do not vary with excitation potential, and similar abundances are obtained from Fe I and Fe II lines. The temperature estimates were then estimated using the wings of H$\alpha$ lines, which are sensitive to small variations in temperature.  We have measured FWHM of telluric and ThAr lines to broaden the synthetic spectra using the Gaussian profile for the resolution(R~30000) of HESP. The logg estimated from the FeI/FeII lines was assumed for the calculation of the line profile for the Balmer line analysis.  The corrections for the non-LTE effects in the estimations of effective temperature from Balmer lines was also incorporated in the adopted values. The color temperatures for the stars were also derived to check for consistency. The different estimates of $T_{\rm eff}$ are listed in table 2. Similarly, $\log g$ is determined by spectral fitting of the wings of the Mg I triplet in the 5173\,{\AA} region. Examples of the fitting for the H$\alpha$ and Mg wings are shown in Figure 1. Independent estimates of $logg$ were carried out by other methods as listed in table 3 - (i) Ionization equilibrium method using Fe I and Fe II abundances, (ii) using the parallax from Gaia as described below. Surface gravity log\,{g}  is  calculated using the relation \\ log(g/g$_{\odot}$) = log(M/M$_{\odot}$) + 4log(T$_{eff}$/T$_{eff\odot}$) + 0.4(M$_{bol}$ $-$ M$_{bol\odot}$)\\
The V magnitudes have been taken from  SIMBAD and the parallaxes are taken from \textit{Gaia} \footnote{https://gea.esac.esa.int/archive/} whenever possible. We have used the evolutionary tracks \footnote{http://pleiadi.pd.astro.it/} to estimate the mass of the stars and are found to be close to 0.8 M$_{\odot}$ for metal-poor stars. The finally adopted stellar parameters are listed in table 4 where we have taken the estimations of $logg$ using Mg wings due to their sensitivity for small changes in $logg$ and $T_{\rm eff}$ using Fe I lines as large number of clean Fe I lines could be measured for every star. The parameters were consistent within the typical uncertainties of $\sim$150 K for temperature and 0.25 dex for $\log g$. However, we also report a discrepancy between the color temperatures and spectroscopic temperatures for the three stars SDSS J1341+4741, SDSS J1725+4202 and SDSS J1933+4524 as shown in table 2. 

Errors in the derived abundances primarily depend on the signal-to-noise ratio (SNR) of the observed spectra and deviations in the values of the adopted stellar parameters. We have used the relation given by \cite{cayrel1988} to calculate the uncertainty in the abundances due to the SNR. The typical uncertainties in the derived stellar parameters are taken to be $\sim$150 K for temperature and 0.25 dex for $\log g$.

\begin{figure}
\centering
\includegraphics[width=1.0\columnwidth]{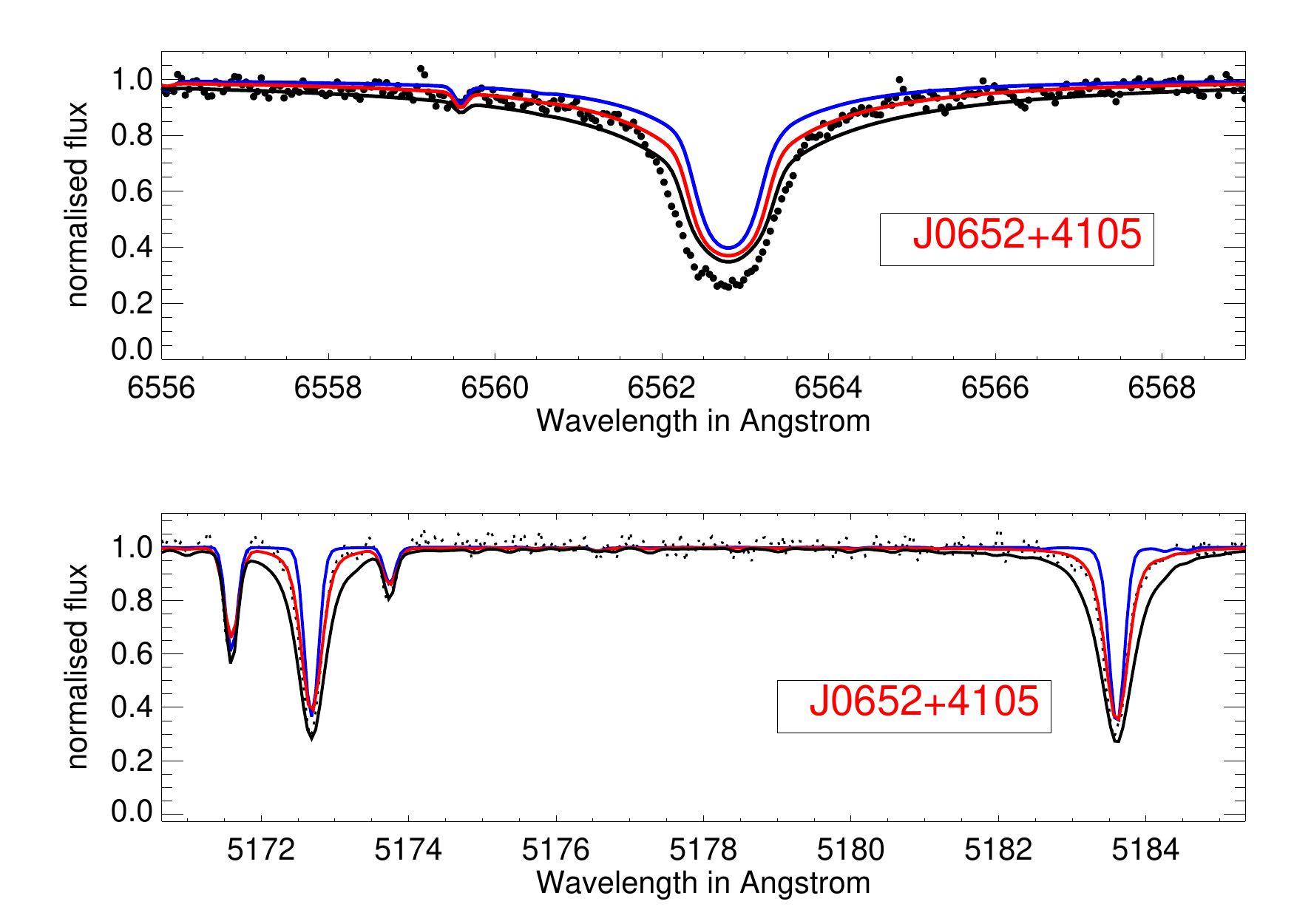}
\caption{The determination of stellar parameters. The H$\alpha$ feature is shown in the top panel and the wavelength range shown here for the fitting is 6556 to 6572 \AA  while the Mg triplet region is shown in the bottom panel for the wavelength range 5170 to 5186 \AA. In the top panel, the red line indicates the best fit to the wings, while black-filled dots indicate the observed spectra. A wavelength range of 10 \AA ~on either side of the core is used for fitting the line wing profile. The black and blue lines indicate deviations from the adopted temperature by 150 K. In the bottom panel, the red line denotes the best fit, while the black and blue lines show the deviation from the adopted $logg$ by 0.50 dex.}
\end{figure}

\begin{table}
\begin{center}
\caption{Estimates of Effective Temperature (K) for the Program Stars}
\begin{tabular}{crrrrrrrrrrr}
\hline\hline
Object  &H-$\alpha$ &Fe~I &$V-K$ &$J-H$ &$J-K$\\
\hline
SDSS J0024+3203 &5700 &5700 &5737 &5672 &5875\\
SDSS J0315+2123 &5450 &5400 &5570 &5287 &5264\\
SDSS J0643+5934 &4800 &4900 &4618 &4843 &4801\\
SDSS J0646+4116 &5100 &5150 &5065 &5179 &5144\\
SDSS J0652+4105 &4900 &5000 &5060 &5108 &5060\\
SDSS J1024+4151 &4800 &4800 &4655 &4782 &4823\\
SDSS J1146+2343 &5200 &5100 &5825 &5273 &5365\\
SDSS J1341+4741 &5450 &5450 &5927 &5438 &5749\\
SDSS J1725+4202 &5300 &5400 &6274 &5803 &6012\\
SDSS J1933+4524 &5850 &5800 &6249 &6038 &6249\\
SDSS J1937+5024 &4950 &4800 &4738 &4702 &4908\\
SDSS J1953+4222 &5900 &6000 &6136 &5847 &5874\\
\hline
\end{tabular}
\end{center}
\end{table}

\begin{table}
\begin{center}
\caption{Different estimates of $\log g$ for the Program Stars}
\begin{tabular}{crrrrrrrrrrr}
\hline\hline
Object & $FeI/FeII$ & Mg-wings & $Gaia$ parallax\\
\hline
SDSS J0024+3203 &3.80 &3.75 &3.94 \\
SDSS J0315+2123 &4.50 &4.50 &4.27\\
SDSS J0643+5934 &2.25 &2.50 &2.28\\
SDSS J0646+4116 &2.25 &2.25 &2.46\\
SDSS J0652+4105 &2.75 &2.50 &2.49\\
SDSS J1024+4151 &2.50 &2.50 &2.38\\
SDSS J1146+2343 &3.10 &3.00 &3.12\\
SDSS J1341+4741 &2.60 &2.50 &2.97\\
SDSS J1725+4202 &3.75 &3.50 &3.90\\
SDSS J1933+4524 &4.40 &4.50 &4.32\\
SDSS J1937+5024 &1.50 &1.50 &1.97\\
SDSS J1953+4222 &3.75 &4.00 &3.97\\
\hline
\end{tabular}
\end{center}
\end{table}

\begin{table}
\begin{center}
\caption{Adopted stellar parameters for the Program Stars}
\begin{tabular}{crrrrrrrrrrr}
\hline\hline
Object & $T_{\rm eff}$ (K) & $\log g$ (cgs) & $\xi$ &[Fe/H] &$A$(Li)\\
\hline
SDSS J0024+3203 &5700 &3.75 &1.50 &$-$2.45 &2.00\\
SDSS J0315+2123 &5400 &4.50 &1.00 &$-$2.30 &1.80\\
SDSS J0643+5934 &4900 &2.50 &1.50 &$-$2.90 &0.80\\
SDSS J0646+4116 &5150 &2.25 &1.50 &$-$1.90 &1.00\\
SDSS J0652+4105 &5000 &2.50 &1.50 &$-$2.56 &1.75\\
SDSS J1024+4151 &4800 &2.50 &1.50 &$-$2.25 &1.05\\
SDSS J1146+2343 &5100 &3.00 &1.00 &$-$2.60 &1.15\\
SDSS J1341+4741 &5450 &2.50 &1.80 &$-$3.20 &1.95\\
SDSS J1725+4202 &5400 &3.50 &1.20 &$-$2.50 &1.90\\
SDSS J1933+4524 &5800 &4.50 &1.80 &$-$1.80 &2.25\\
SDSS J1937+5024 &4800 &1.50 &1.50 &$-$2.20 &1.00\\
SDSS J1953+4222 &6000 &4.00 &1.75 &$-$2.25 &2.05\\
\hline
\end{tabular}
\end{center}
\end{table}

\section{Abundances}

The results of our abundance analysis for 8 of the program stars are provided in Tables 4-11 below.  Here we discuss details of this analysis for various classes of elements.

\subsection{Lithium}

\begin{figure}
\centering
\includegraphics[width=1.0\columnwidth]{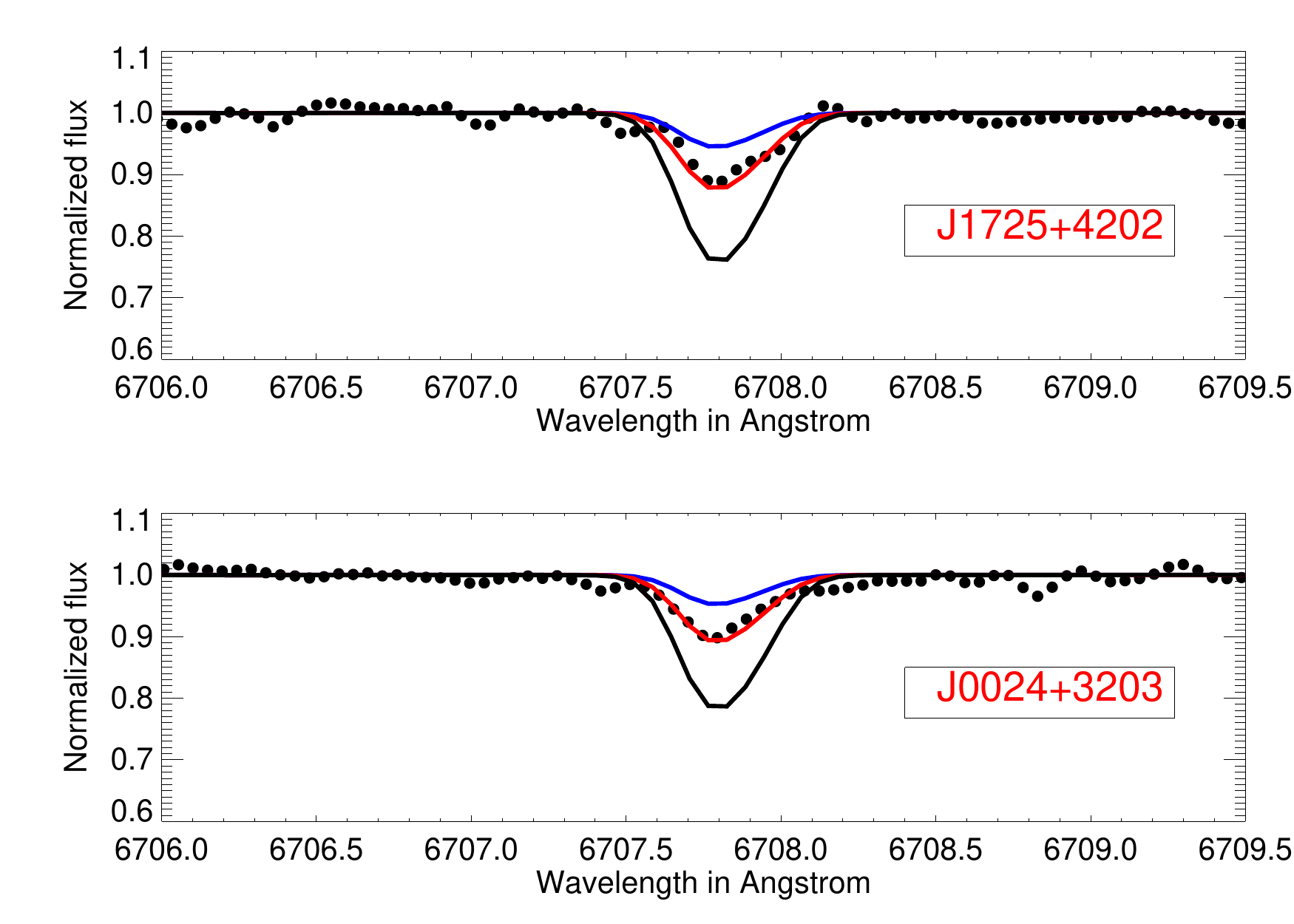}
\caption{The synthesis of the Li line. Red marks the best-fit synthetic spectrum, while black-filled circles indicate the observed spectra. The black and blue lines show the synthetic spectra for deviations in the Li abundance of 0.50 dex from the best fit.} 
\label{fig:lith-line}
\end{figure}

Lithium abundances were derived from the strong absorption features at 6707.76\,{\AA} and 6707.98\,{\AA}, using the method of spectrum synthesis. The continuum level for the observed spectra is estimated locally around the Li doublet. The observed spectra were fit iteratively with the synthetic spectra for different values of Li abundance, and the best fit was adopted for each star, keeping the Li abundances as the only free parameter in the synthesis. Examples of the spectral synthesis for Li are shown in Figure~\ref{fig:lith-line}.
 
Errors in the abundance analysis of Li primarily originate from uncertainties in estimates of effective temperature. A difference of $\sim$ 150\,K is found to alter the Li abundance by 0.14 dex, on average. For the determination of the abundances of neutral species like Li~I, uncertainties in surface gravity play a minimal role.
 
 \subsection{The light and ${\alpha}$-elements}

Abundances of carbon could be derived for all of the program stars based on the molecular CH $G$-band in the 4315\,{\AA} region. Most of the stars have low C, the range varying between [C/Fe] = $-$0.53 to [C/Fe] = +0.22. Corrections to the measured carbon abundances due to the evolutionary effects were computed based on \cite{placco2014} and were found to be minimal (0.0 - 0.01 dex) for the program stars which have been incorporated in the final reported C abundances in the tables. The poor signal-to-noise (SNR) in the region of the CN band at 3883\,{\AA} did not allow for precise abundances for N, while the region containing the O lines at 6300\,{\AA} and 6363\,{\AA} were too weak and dominated by telluric contamination, which prevented a meaningful derivation of O abundances for most of the stars. Oxygen could be derived for \ten and was found to be enhanced, [O/Fe] = +1.56.

Among the odd-Z elements, Na and Al could be detected and measured for all of the program stars. The Na abundances were determined using the D1 and D2 lines at 5890\,{\AA} and 5896\,{\AA}; the Al abundances were measured based on the resonance line at 3961.5\,{\AA}. NLTE corrections for both the elements, based on \cite{andrievskyal, andrievskyna}, were also implemented, as reported in Tables 4-11. The mean abundances are shown in comparison to samples from \citet{cayrel2004} and \citet{cohen2004} in Figure \ref{fig:cohen}. In this study, Si abundances could be derived for 5 stars out of the total sample of 11 stars. The abundances are mostly based on the line at 410.29 nm (and also 390.55 nm in a few stars with high blending from the CH line) which falls at the wings of the $H-\delta$ line apart from having a very poor signal-to-noise ratio. However, the average Si abundances in our study agree well with \cite{cayrel2004} as demonstrated in figure 4 but the Si abundances for \cite{cohen2004} are lower due to the spectral syntheses to determine the abundance of Si in the C-rich stars. These yield abundances of Si are substantially lower than those obtained with the standard analysis and are indicated in Tables 4–7 and the lower panel of Figure 4 in \cite{cohen2004}. Similarly, carbon is also higher primarily due to the evolutionary effects. The spectral synthesis is known to yield more accurate abundances for the weaker lines and low SNR of the spectra and a mixture of both methods have been used in this study. The uncertainties in the abundances for each element have also been indicated in figure 4.

\begin{figure}
\centering
\includegraphics[width=1\columnwidth]{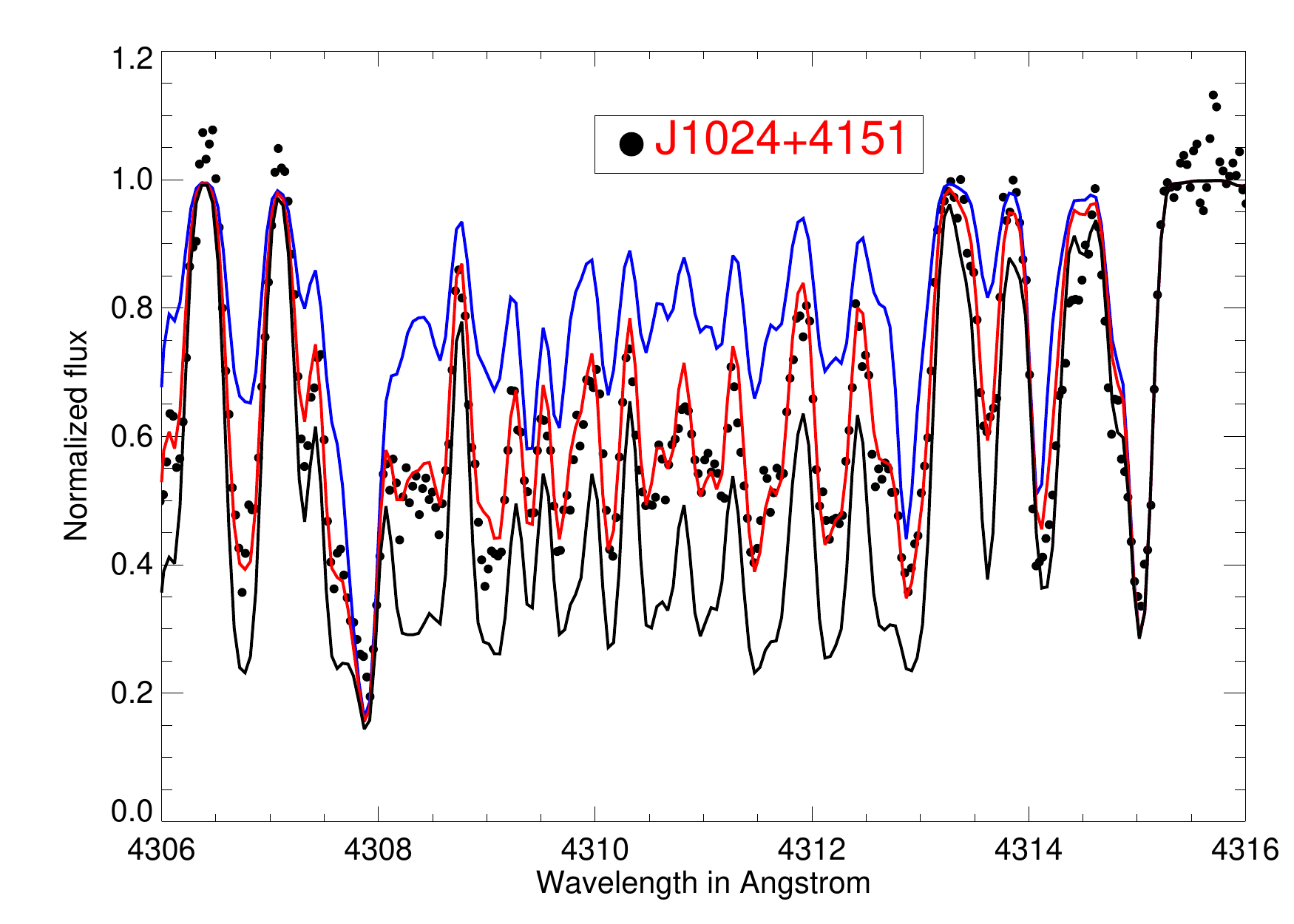}
\caption{An example of spectral fitting for the molecular CH $G$-band region. The black-filled circles denote the observed spectra while the coloured lines indicate the synthetic spectra. Red shows the best spectral fit, while black and blue lines indicate the spectra for C abundance deviating by $\pm$0.50 dex from the best fit.}
\end{figure}

The $\alpha$-elements are produced in different astrophysical sites, such as the hydrostatic burning phases in the shells of massive stars, oxygen and neon burning in Type II supernovae, and hypernovae. Among the $\alpha$-elements, Mg, Ca, and Ti abundances could be derived for all of the program stars, but meaningful Si abundances could only be derived for a few stars due to the poor SNR towards the blue end of the spectra. Several lines of Mg, Ca, and Ti could be detected in the spectra; the method of equivalent widths was employed to determine the abundances for the stronger lines, while spectral synthesis was used to determine abundances from the weaker and blended features.

Uncertainties in the derived abundances primarily depend on the signal-to-noise ratio (SNR) of the observed spectrum and deviations in the values of the adopted stellar parameters. We have used the relation given by \cite{cayrel1988} to calculate the uncertainty in the abundances due to the SNR. Uncertainties due to possible temperature and $\log (g)$ deviations were derived using two different model spectra, the first differing in temperature by $\sim$150 K and the second one deviating in $\log (g)$ by 0.25 dex. The final values of the abundance errors were obtained by adding the uncertainties arising from all three sources in quadrature. However, the errors in the relative abundance ratios are less sensitive to the errors in the model parameters and  mainly depend on the SNR.

 \subsection{The Fe-peak elements}

\begin{figure}
\centering
\includegraphics[width=1.0\columnwidth]{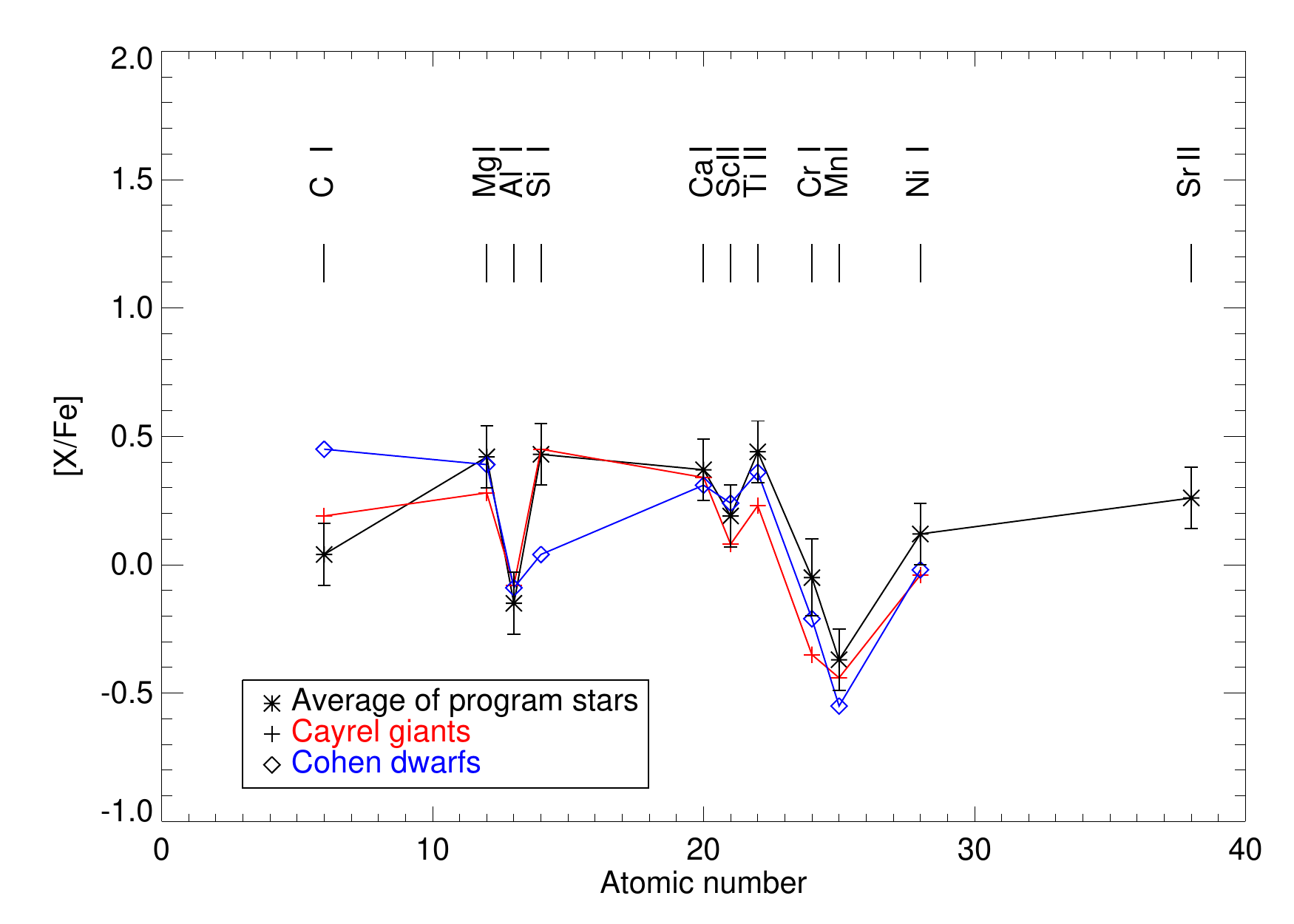}
\caption{The nucleosynthesis pattern for the program stars, compared to the samples from \citet{cayrel2004} and \citet{cohen2004}. The error bars corresponding to the uncertainties in the abundances for each element have been demonstrated in black lines.}
\label{fig:cohen}
\end{figure}

The Fe-peak elements (Sc, Fe, Cr, Mn, Co, Ni, and Zn) are synthesized during complete and incomplete Si burning phases in pre-supernovae, as well as during the explosive phase of a Type II supernova. Iron abundances were derived on the basis of several Fe I and Fe II lines; a difference of 0.25 dex was noted, which is in agreement with other analyses of metal-poor stars. The iron abundance of the program stars varies from [Fe/H] = $-1.80$ to $-3.20$, with a mean value of [Fe/H] = $-2.40$. The abundances of Cr were measured from Cr I lines, which are known to suffer from strong NLTE effects \citep{lai2008, bonifacio2009}; a mean difference of 0.35 dex was obtained between the Cr I and Cr II lines in the current sample. Manganese abundances were primarily measured using the resonance triplet near 4030\,{\AA}, but the poor quality of the spectra in that region led to larger errors. The NLTE corrections for the Mn triplet region increase with decreases in metallicity \citep{bergemann2008,bergemann2019}. The mean value for the present sample is [Mn/Fe] = $-0.37$. \zerosix is very strongly depleted in Mn, with [Mn/Fe] = $-1.00$. Cobalt is a product of complete Si burning, and it tracks the iron content of the star, with the expected scatter due to observational uncertainties. The mean abundance of Co for the present sample is [Co/Fe] = +0.01. The nucleosynthesis pattern for the program stars, in comparison to the mean abundances of giant stars from \cite{cayrel2004} and dwarf stars from \cite{cohen2004}, is shown in Figure~\ref{fig:cohen}. As seen in the figure, the derived abundances agree well with the mean abundances from these samples.

\subsection{The n-capture elements}

Out of the several neutron-capture elements that could be measured in our spectra, abundances of Sr and Ba could be derived for all of the program stars by the method of spectral synthesis. Both the lines at 4077\,{\AA} and 4215\,{\AA} were used to derive the Sr abundances, while the line at 4554\,{\AA} was used to derive the Ba abundance. The other strong Ba line at 4934\,{\AA} was avoided, as analysis of this line is extremely difficult, and yields large errors due to Fe blends found in the wings of this line \citep{gallagher2010}. The average abundances for Sr and Ba for the present sample are [Sr/Fe] = +0.26 and [Ba/Fe] = +0.25. However, several additional n-capture elements could be detected in \sixfiftytwo, which exhibits a large and uniform enhancement of n-capture elements. The first n-capture peak species Sr, Y, and Zr could be detected among the lighter n-capture elements, with an average value of [ls/Fe] = +0.68. Among the heavier n-capture elements, Ba, La, Nd, and Eu could be measured, with an average value of [hs/Fe] = +0.90. The values of [hs/ls]= +0.22 and [Ba/Eu] = $-0.23$ indicates that \sixfiftytwo could have received contributions from both the $r$-process and $s$-process. However, the strong presence of Li, with $A$(Li) = 1.75, along with a low abundance of carbon ([C/Fe]= $-0.12$, corrected for evolutionary effects) rules out the possibility of $s$-process via mass transfer from a companion binary star or winds from a massive star \citep{susmitha2021}. Hence, the origin and evolution of \sixfiftytwo is more likely to be of $r$-process origin. The possibility of $i$-process origin \citep{hampel2016,den2017} needs to be further explored as well.

 \subsection{Kinematics}

To obtain the stars' kinematics, we required distances, proper motions, and radial velocities.  Fortunately, Gaia provides a unique opportunity to estimate the proper motions of the stars with high precision, which we have adopted from Gaia-EDR3 \citep{gaia_edr3}. Distances are obtained from the \citealt{bailerjones2021} catalogue, where these authors have estimated the distances from Gaia-EDR3 parallaxes using probabilistic methods. The radial velocities were derived from the observed spectra, as listed in Table 1. We used the Astropy module to convert into the Galactocentric system and determined Galactocentric distances (X, Y, Z) and velocities ($V_x$,$V_y$, $V_z$), where the XY plane is the Galactic plane. For this conversion, we considered the Galactocentric distance for the Sun to be 8.2 kpc \citep{dehnenbinney98,mcmillan2017}, the distance of the Sun from the Galactic plane as 0 kpc, and the Solar velocity components to be (12.9, 245.6, 7.78) km/s \citep{meingast21}.

We also derived the orbital characteristics of the observed program stars \citep{pinto2021} . For our computation, we have used $r$ to denote the Galactocentric distance. The parameter $V_r$ is the velocity component along $R$, while $V_z$ is the vertical component of the velocity of the stars. The parameter $L_z$ represents the $z$ component of angular momentum, and $L_{\perp}$ denotes the perpendicular component of angular momentum. The parameter $V_{\phi}$ corresponds to the azimuthal velocity and is given by $L_z/R$. All of the computed velocities and angular momenta for the program stars are listed in Table 3. The velocities are listed in km/s, whereas the angular momenta are listed in multiples of 10$^2$ kpc km/s. Following \cite{dimatteo2020}, assuming a clockwise rotation of the disc, a negative value of $V_{\phi}$ represents prograde motion, whereas a positive value of $V_{\phi}$ represents retrograde motion. As shown in Figure 5, the dotted black line separates prograde motions from retrograde motions.  Our program stars are evenly distributed in both regions; four stars towards the bottom left of the diagram likely belong to the disc population. \sixfiftytwo, marked in red, also belongs to this group. The other stars on the Spite plateau from the literature are shown with green open circles. The blue semi-circle shows the expected location of the disc stars in this plane.

\begin{table*}
\tabcolsep6.0pt $ $
\begin{center}$ $
\caption{Kinematics for the Program Stars}
\begin{tabular}{crrrrrrrrrr}
\hline\hline
Object  &X  &Y  &Z  & $V_x$  & $V_y$  & $V_z$  & $V_R$  & $L_z$  & $V_{\phi}$  & $L_{\perp}$\\
        & (kpc) & (kpc) & (kpc) & (km/s) & (km/s) & (km/s) & (km/s) & (10$^2$ kpc km/s) & (km/s) & (10$^2$ kpc km/s)    \\   
        \hline

SDSS J0024+3203 &$-$8.19  &0.19 &$-$0.12  &292.7 &$-$83.0  &152.4  &$-$294.6  &6.2  & 76.0  &12.12 \\
SDSS J0315+2123  &$-$8.38  &0.09  &$-$0.17  &27.2  &158.4  &35.9  &$-$25.5  &$-$13.3  &$-$158.7  &2.98 \\
SDSS J0643+5934  &$-$9.00  &0.56  &0.56  &$-$96.5  &131.8  &13.2  &104.3  &$-$11.8  &$-$125.8  &0.96 \\
SDSS J0646+4116  &$-$9.14  &0.09  &0.31  &216.6  &$-$71.7  &$-$208.3  &$-$217.4  &6.3  &69.4  &18.36 \\
SDSS J0652+4105  &$-$9.21  &0.09  &0.35  &$-$91.0  &224.5  &12.6  &93.2  &$-$20.6  &$-$223.6  &1.15 \\
SDSS J1024+4151  &$-$9.06  &0.04  &1.46  &$-$278.6  &$-$5.6  &56.5  &278.5  &0.6  &7.1  &1.05 \\
SDSS J1341+4741  &$-$8.13  &0.16  &0.40  &202.4  &$-$79.2  &$-$49.9  &$-$204.0  &6.1  &75.1  &3.25 \\
SDSS J1725+4202  &$-$8.02  &0.17  &0.12  &$-$218.0  &7.8  &$-$7.1  &218.1  &$-$0.2  &$-$3.1  &0.84 \\
SDSS J1933+4524  &$-$8.05  &0.23  &0.05  &273.3  &340.2  &72.6  &$-$263.2  &$-$28.0  &$-$348.1  &5.98 \\
SDSS J1937+5024  &$-$7.96  &1.12  &0.27  &157.4  &58.6  &$-$81.0  &$-$147.6  &$-$6.4  &$-$80.1  &6.10 \\
SDSS J1953+4222  &$-$8.07  &0.12  &0.02  &$-$121.0  &$-$45.9  &33.2  &120.3  &3.8  &47.8  &2.65 \\
\hline
\end{tabular}
\end{center}
\end{table*}

  \begin{figure}
\centering
\includegraphics[width=1.0\columnwidth]{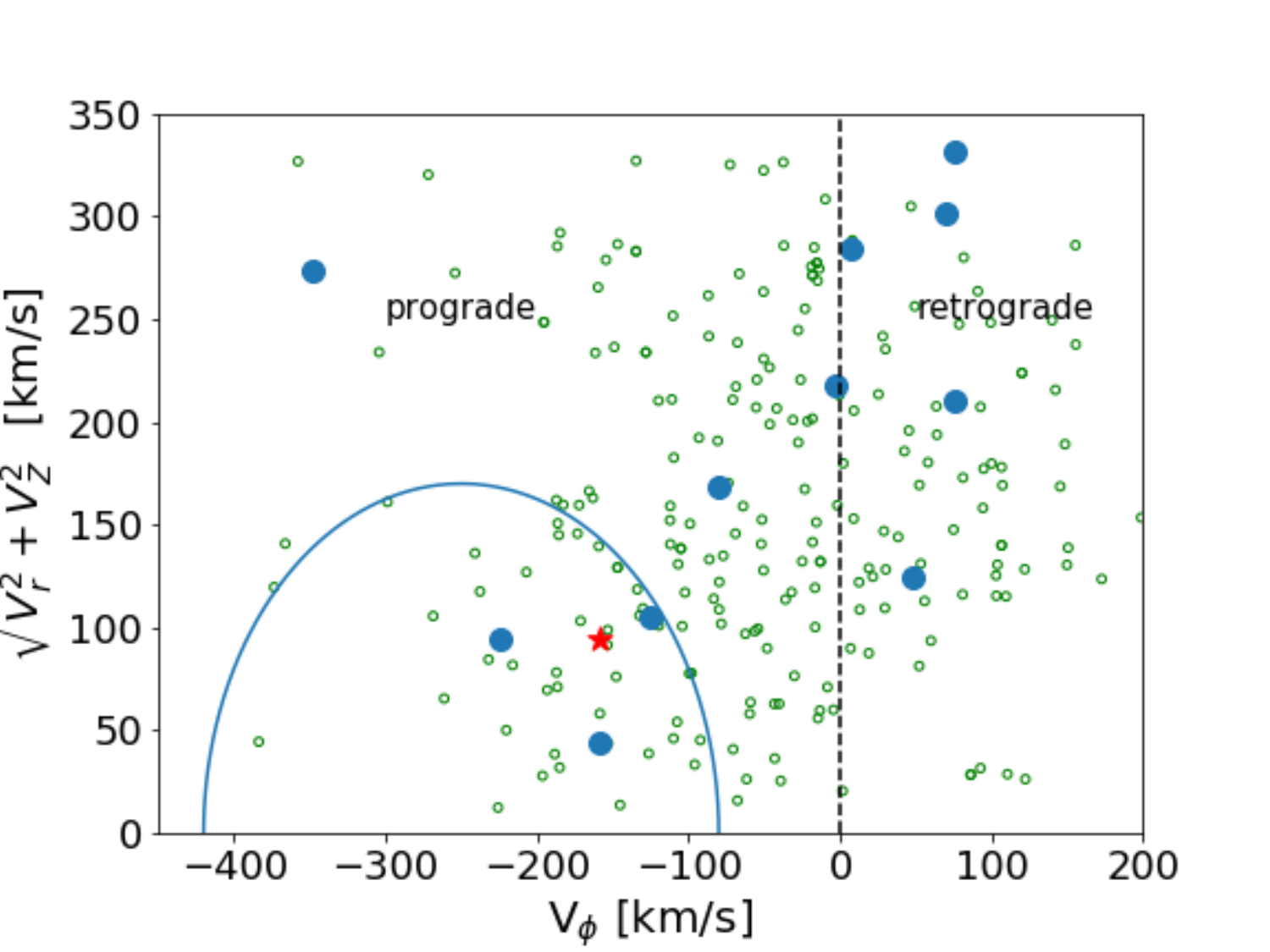}
\caption{Toomre diagram exhibiting the orbital characteristics of the program stars. The dotted line indicates the split between prograde and retrograde motions. The blue-filled circles represent the program stars, while the red-filled star indicates \sixfiftytwo. The blue semi-circle shows the expected location of the disc stars in this diagram. The green open circles represent stars in the Spite plateau from the literature.} 
\end{figure}

\begin{figure*}
\centering
\includegraphics[width=1.8\columnwidth]{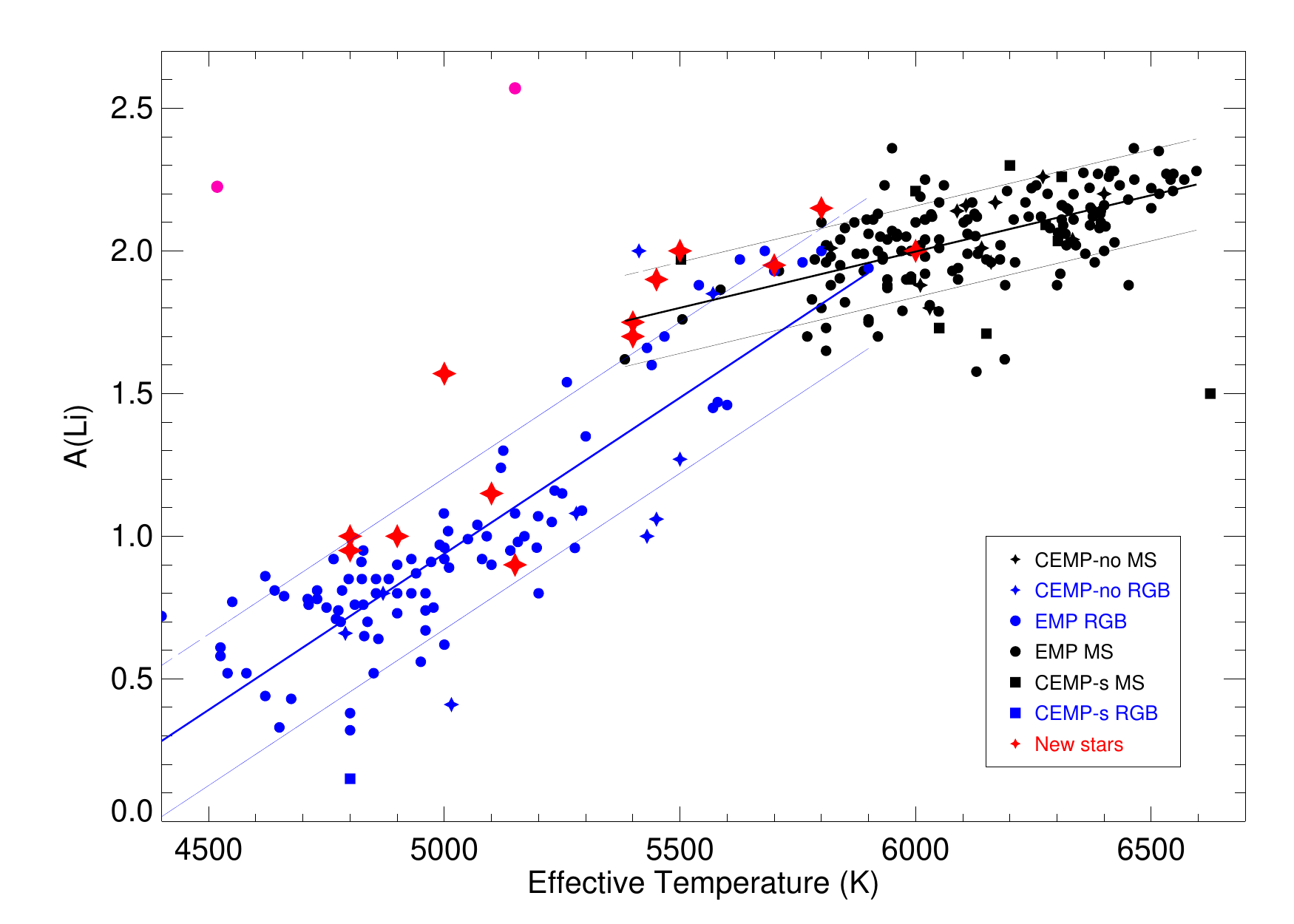}
\caption{The distribution  $A$(Li), as a function of $T_{\rm eff}$, for the different stellar families indicated in the figure. The program stars are marked by large red diamonds. The best fit for the giants (in blue) and dwarfs (in black) are shown by solid lines along with the 1$\sigma$ regions for each fit. The slope and $\sigma$ of the ﬁt for the giants are 0.0012 dex/K and 0.19 dex/K, respectively, while for the dwarfs they are 0.0005 dex/K and 0.12 dex/K, respectively.}
\end{figure*}

\section{Results and Discussion}

\subsection{Lithium distribution in the metal-poor regime}


We have demonstrated the distribution of Li abundances as a function of temperature for different stellar families in figure 6. The stars have been further categorized into giants and dwarfs for the EMP, CEMP-no, and CEMP-$s$ classes. The literature data is compiled from the SAGA database \citep{sudasaga}, and the program stars with detections of Li are marked in red diamonds. As noted by several studies on Li abundances in metal-poor stars \citep{spitenspite,bonifacio2007,litospite}, the plateau is observed for warmer dwarf stars with $T_{\rm eff}$ $>$ 5800\,K. However, the scatter tends to increase as temperature decreases from 6500\,K to 5700\,K. 
The identical distribution of Li across the EMP and CEMP-no stars indicates the ISM to be well-mixed during the epochs of their formation. Hence, Li could not be used as a yardstick to differentiate between these different stellar populations. However, Li is often depleted for CEMP-s stars but they are not considered for deriving the fits in figure 6 as the Li in these stars is often accreted through mass transfer from a companion AGB star. Mass transfer from a low-mass AGB would produce large amounts of C and deplete Li, along with the production of $s$-process-enhanced material. There are models in which AGB stars could produce Li through the Cameron-Fowler mechanism \citep{cameronfowler1971}. Through this mechanism, the outer convective envelope comes in contact with the H-burning shell where \textsuperscript{3}He is being produced by proton-proton reactions. The \textsuperscript{3}He is burned to \textsuperscript{7}Be via \textsuperscript{3}He($\alpha$,$\gamma$)\textsuperscript{7}Be under convective conditions. The \textsuperscript{7}Be is then swept up to the stellar surface and decays to \textsuperscript{7}Li by electron capture.. Only three of our sample stars with measurable Li are MSTO stars, which is not adequate to test the consistency of the slope for our sample. Black and blue dots are used in this figure for the literature sample to homogeneously differentiate between and dwarfs (black) and giants (blue) for all classes of stars. The two outliers among the EMP stars shown in pink filled circles are CS22893-010 \citep{roederer2014} and C1012254-203007 \citep{ruchti2011}.

We have demonstrated the trends for Li abundances in RGB and MS stars \textit{(stars with logg values greater than 4 are considered to be MS stars)} with temperature; definite trends are present.  A strong correlation is obtained for RGB stars, with a Pearson correlation coefficient of 0.89 while the dwarf MS stars exhibit a weaker correlation coefficient of 0.60.  The probability of no-correlation for the RGB and dwarf MS stars are less than  10\textsuperscript{-5}. The best fits for these two populations are shown in blue and black solid lines, respectively in Figure 6. About 85$\%$ of the stars fall within the 1$\sigma$ width of the best fit shown in the solid lines in the plot. The errors for the Li abundances are taken to be 0.05 dex, while that for temperature is taken to be 150 K. The empirical relation governing the best fit for the trends of A(Li) with $T_{\rm eff}$ in the giant stars and dwarf stars are given in the relations $(i)$ and $(ii)$ below, respectively:

\begin{equation*}
\begin{split}
A{\rm (Li)} & = 0.00108\, T_{\rm eff} - 4.524~~~ (i)\\
A{\rm (Li)} & = 0.00037\, T_{\rm eff} - 0.392-~~ (ii)
\end{split}
\end{equation*}

Figure 7 shows the trend of Li abundance with metallicity. The cosmological value of Li and the observed value of the Spite plateau are shown in black lines. A small slope can be seen, and the scatter tends to increase for MS stars as metallicity decreases. The lowest-metallicity stars have lower values of $A$(Li), which reach the Spite plateau at [Fe/H] $> -3.50$, albeit with a large scatter. Our small sample could not yield significant results, but detection of Li in additional stars would provide better opportunities to understand the evolution and (possible) depletion of Li in the early Universe.

\begin{figure*}
\centering
\includegraphics[width=1.8\columnwidth]{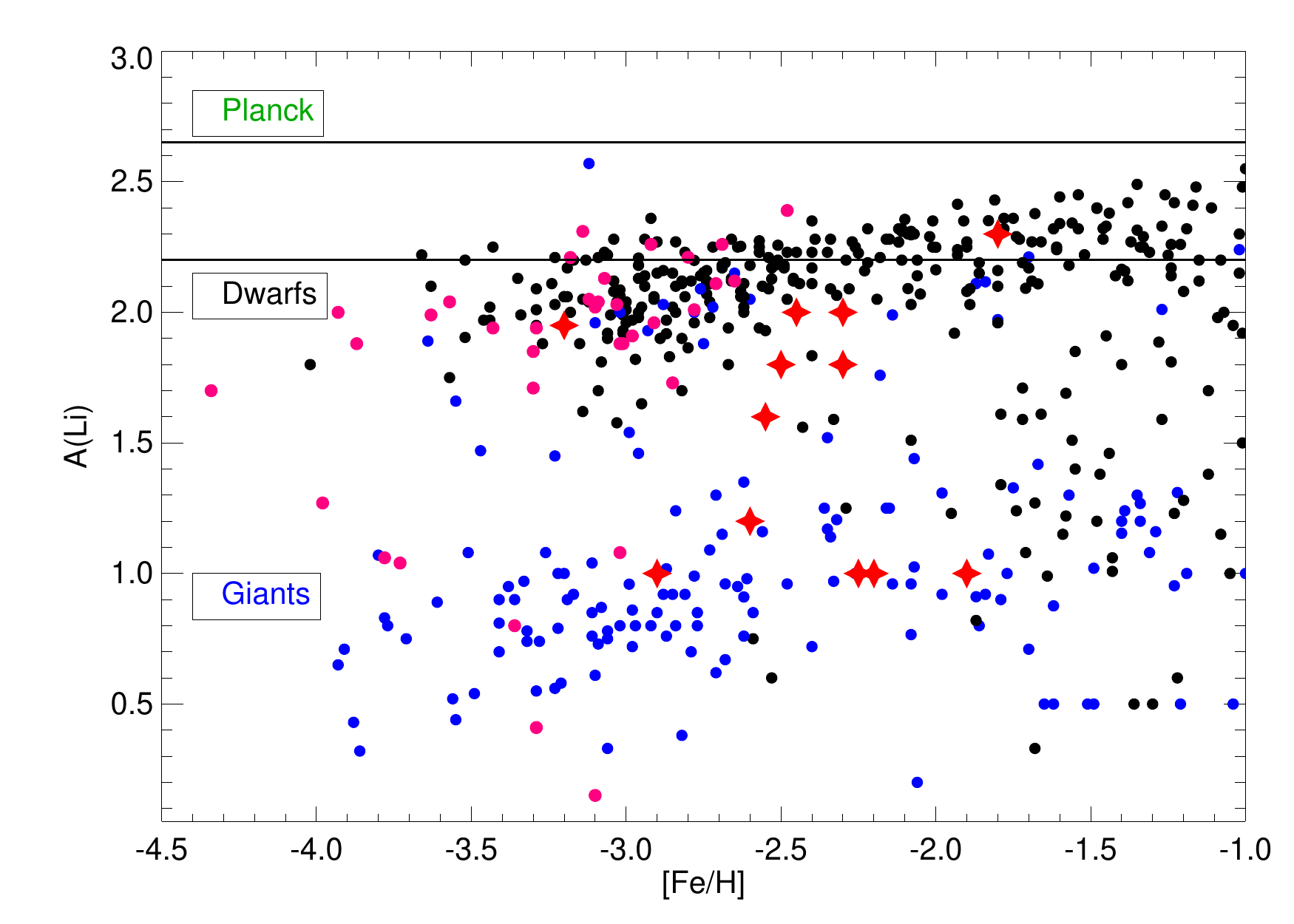}
\caption{The distribution of $A$(Li) as a function of [Fe/H]. Predictions from the Planck mission and the Spite plateau abundances are shown by black solid lines. The dwarf stars are marked in black, while giant stars are marked in blue. CEMP-no stars are shown as red dots. The program stars with Li detections in this study are shown by filled red diamonds.}
\end{figure*}

\begin{figure*}
\centering
\includegraphics[width=1.8\columnwidth]{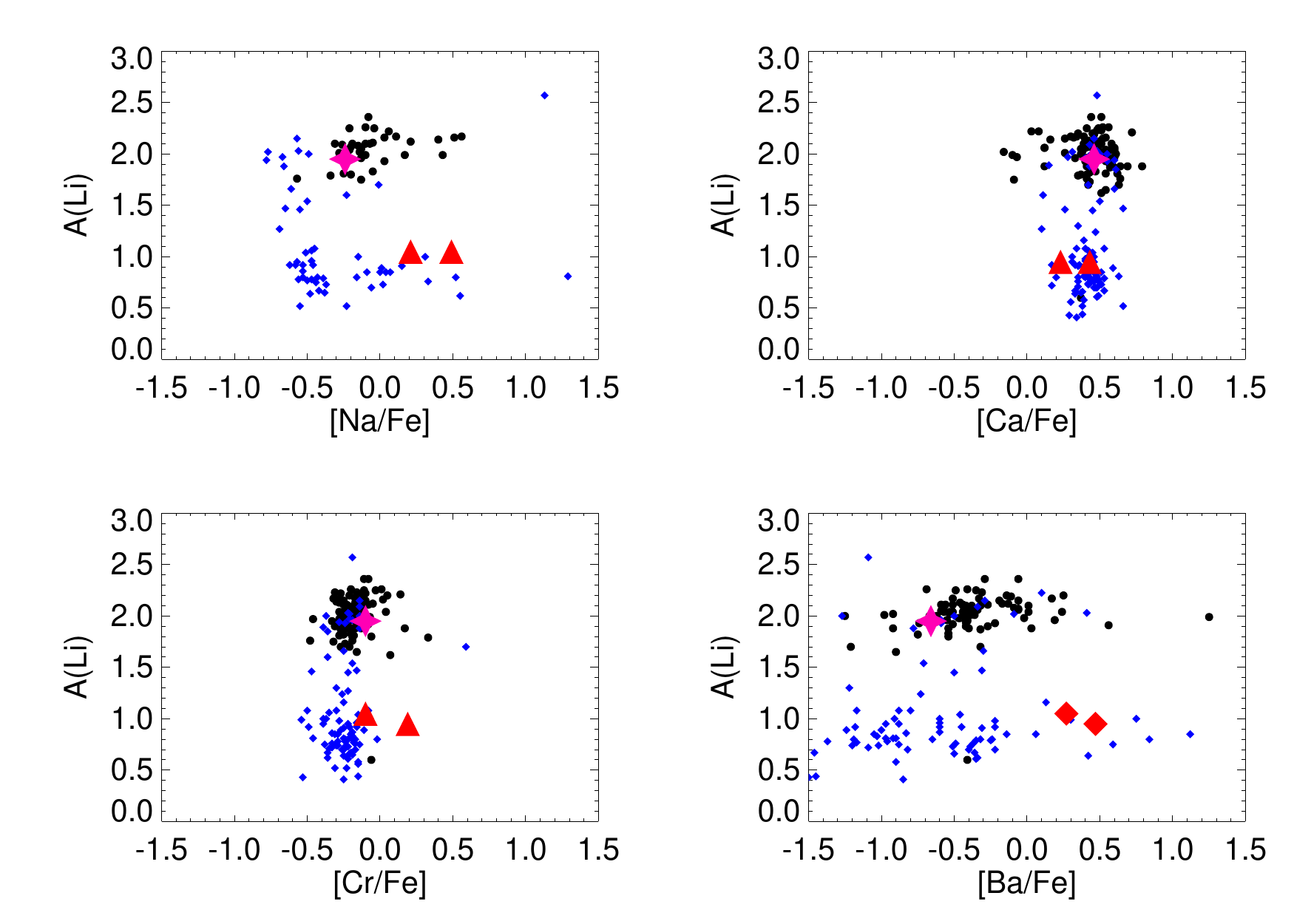}
\caption{  The Li abundance for VMP and EMP stars with detections of Li. The black dots mark the dwarf stars, while blue dots represent the giants. The GC escapees from this study are marked with red-filled triangles, and the CEMP-no star is marked with a pink diamond.}
\end{figure*}

\subsection{Lithium abundances in halo and globular cluster stars}

\citet{pasquini} found a trend for Li abundances with other elements, such as Na, O, and N, for the MSTO stars in NGC 6752. $A$(Li) was found to correlate with [O/Fe] and anti-correlate with [Na/Fe] and [N/Fe]. \citet{bonifacio2007} confirmed the $A$(Li)-[Na/Fe] anti-correlations in 47 Tucanae. However, no such trend was noticed among the halo stars. In Figure 8, the halo stars have been divided into dwarfs (black dots) and giants (blue dots). The depletion of $A$(Li) due to evolutionary mixing can be seen in all four panels. No trends could be seen for $A$(Li) with [Na/Fe] (as reported in a few GCs), [Ca/Fe] (representative of $\alpha$-elements abundances), [Cr/Fe] (an Fe-peak element), and [Ba/Fe] (a n-capture element). We have also marked the abundances of the GC escapees \citep{ban_gce} and the CEMP-no star \citep{bandyopadhyay} with red-filled triangles and pink diamonds respectively in each of the panels.

Lithium could also be detected for both of the GC escapees of this study, \zerosixgc and \nineteengc \citep{ban_gce}, and is found to be normal. Lithium is a fragile element, which is completely destroyed in a temperature range much lower than that required for the operation of the Mg-Al cycle. Thus, the presence of Li in second-generation stars indicates a heavy dilution of the gas processed by p-capture reactions with unprocessed gas that still preserves the standard Population II lithium abundance \citep{dantona2019}. Lithium has been measured in several Galactic GCs \citep{dorazi2015, dantona2019}; the Li abundances exhibit a similar distribution as normal metal-poor halo stars.

\subsection{Li abundance as a function of distance from the Galactic plane}

When combined with stars from the literature, we have 337 stars with Li, [Fe/H], and RV information. In Figure 7, we showed that there is a plateau with a negative slope for the $A$(Li) distribution with [Fe/H]. We wanted to examine if there is any correlation between the distribution of stars with respect to distance from the Galactic plane as a function of the $A$(Li) or [Fe/H] abundances. As the present sample combined with the stars from literature are metal-poor and Population II stars it is expected that they will be found more often in the halo.

\begin{figure}
\centering
\includegraphics[width=1.0\columnwidth]{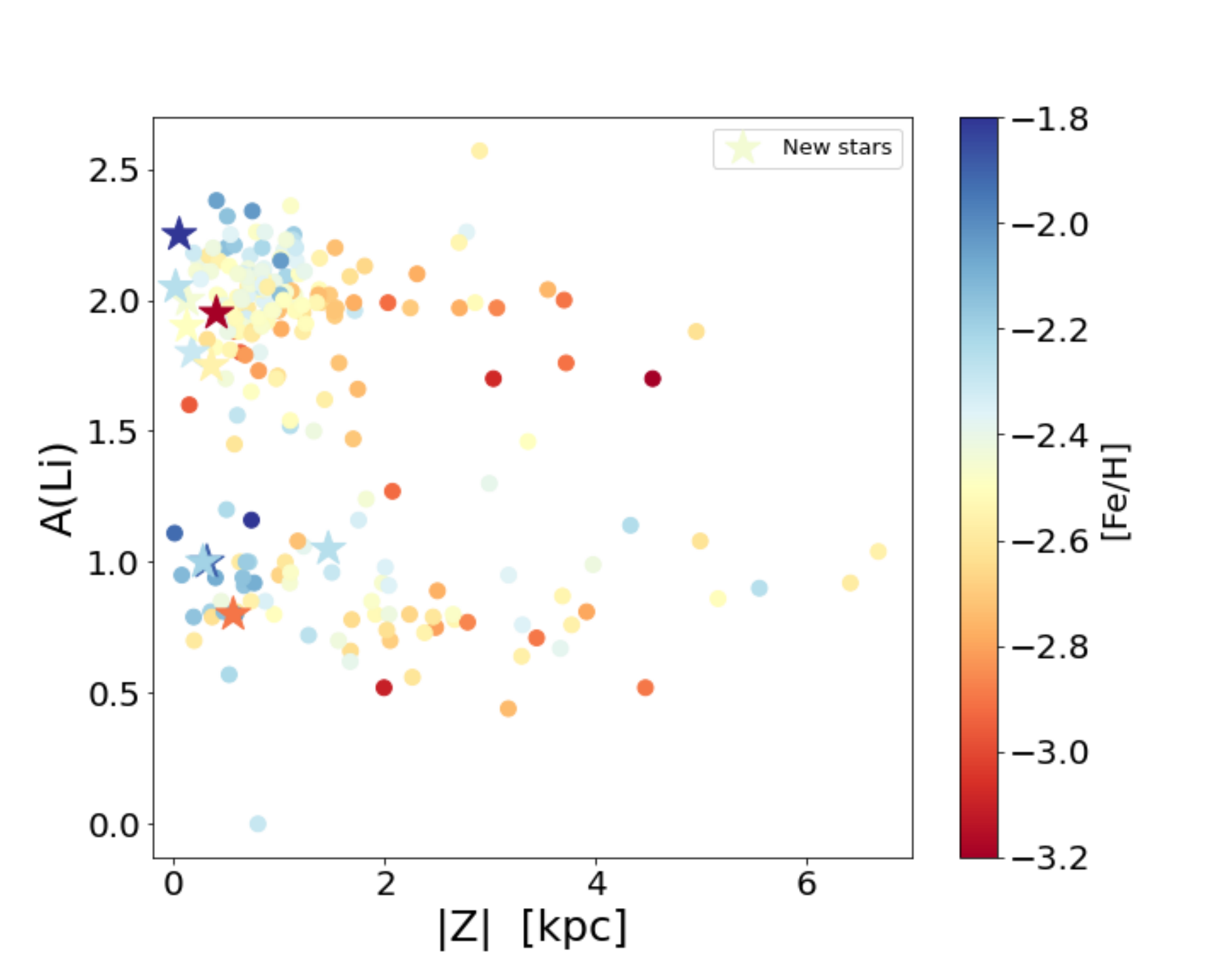}
\caption{Li abundance as a function of distance from the Galactic plane. The colour bar shows the metallicity of the stars. The new stars are marked in filled diamonds.}
\label{li_vs_z}
\end{figure}

Figure 9 shows the relationship between $A$(Li) and absolute distance from the Galactic plane, $|{\rm Z}|$, colour-coded to indicate [Fe/H]. The figure suggests that the Galactic disc ($|{\rm Z}| < 2$ kpc) is mostly populated with relatively metal-rich stars, while a light trend can be noticed between stars' distance to the Galactic plane with their Li abundances. The Li abundances in the stars tend to decrease as their distances from the galactic plane increase. Most Li-rich stars are found to be in or close to the galactic plane. The new sample of stars in this study is demonstrated in filled diamonds with colors corresponding to their metallicity. They are found to populate the region within 2 kpc of the galactic plane.

\subsection{Light-element abundances}

The program stars exhibit the typical odd-even nucleosynthesis pattern exemplified by low Na, high Mg, and low Al \citep{truran1971,umeda2000,hegerandwoosley2002}. The trends for the abundances of odd-Z and $\alpha$-elements are shown in Figure 10. The giants and main-sequence stars exhibit a similar distribution for the light elements. They show the expected enhancement in $\alpha$-elements for the halo stars, with an average [$\alpha$/Fe]= +0.41. The $\alpha$-elements Mg, Ca, and Ti exhibit a consistent over-abundance, but a large scatter is observed in the case of Si. The star \sixfiftytwo (shown in red in Figure 10), apart from being rich in n-capture elements, also exhibits a higher abundance of the odd-Z element Na along with the $\alpha$-elements. The light element abundances for the program stars are found to agree with the previous investigations of metal-poor halo stars as shown in figure 10. The data for the metal-poor halo stars were compiled from the Saga database \citep{sudasaga}. The two GC escapees marked in filled blue diamonds exhibit an elevated abundances of Al as seen in the top right panel. Si could be determined only for 5 out of the 11 stars as shown in the bottom panel.

\begin{figure*}
\centering
\includegraphics[width=1.8\columnwidth]{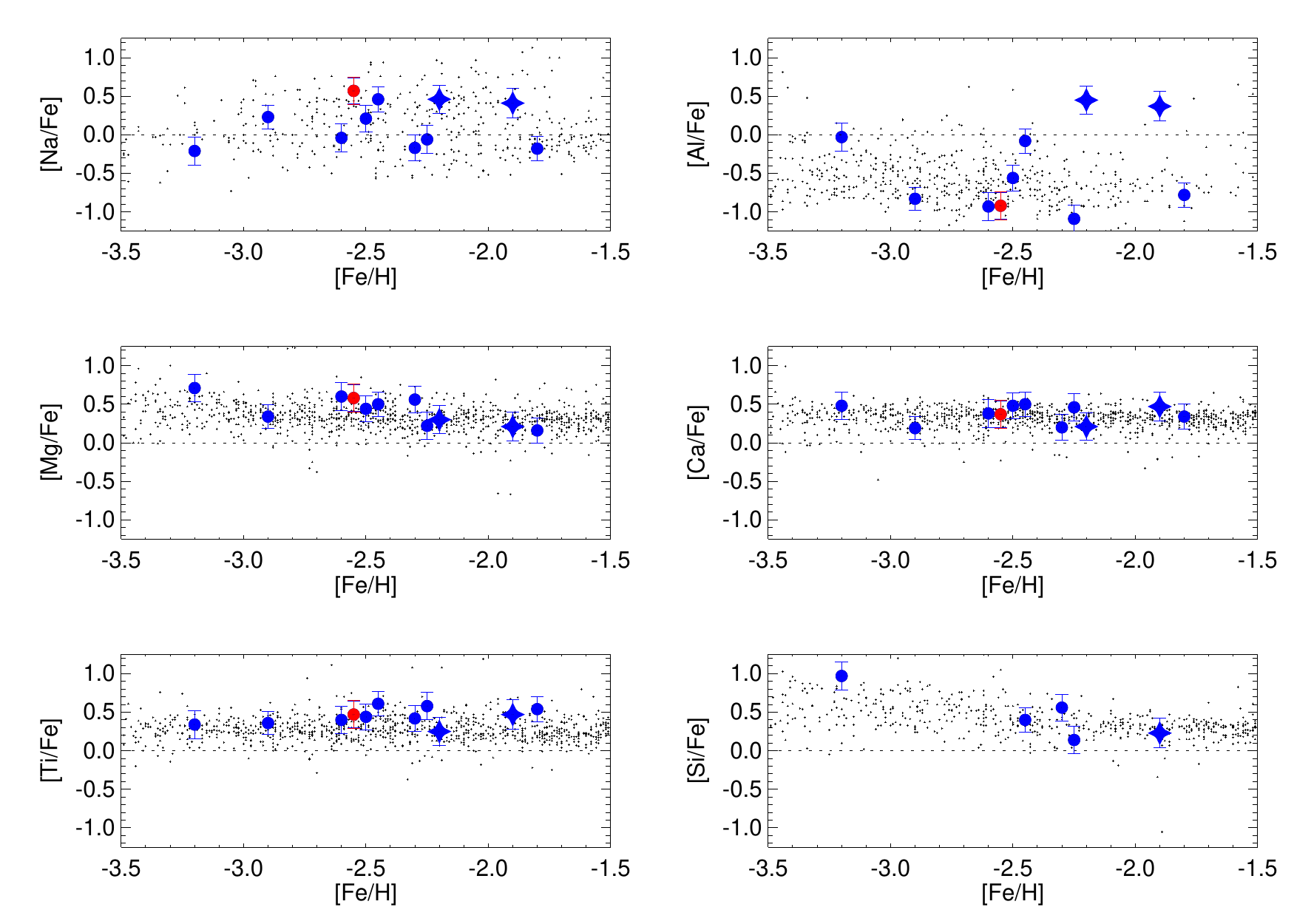}
\caption{The distribution of the light elements Al, Mg, Ca, Ti, and Si as a function of [Fe/H]. The blue-filled circles represent the program stars, while the red-filled circle denotes the abundances for \sixfiftytwo. The GC escapees are shown in blue-filled diamonds. The abundances of other metal-poor stars in literature compiled from \cite{sudasaga} are shown in black dots. The uncertainties are demonstrated by coloured error bars. }
\end{figure*}

\subsection{Heavier-element abundances}

The heavier elements show the expected behaviour with respect to variation in metallicity. The Fe-peak elements appear to closely track the Fe abundances (see Figure 11). Cobalt exhibits the typical decline with respect to increasing metallicity, the Ni and Zn abundances are slightly enhanced, and Mn is mostly depleted, as seen in other VMP stars. The decreasing trend in Cr with metallicity observed in metal-poor stars is due to the NLTE effects on neutral Cr, which varies with metallicity  \citep{bergemanncescutti2010}.  For our sample, Cr varies by 0.3 dex over a 1.4 dex range in metallicity. A large dispersion in Sc abundances was expected from chemical evolution models that include significant contributions from a few supernovae with different masses, but this is not found for the current sample. The well studied trends in abundances for the Fe-peak elements in metal-poor stars are demonstrated in figure 11. The sample size for this study is not adequate to study the variation of individual elements with metallicity over a wide range but the program stars were found to follow the general trend. The large excess in Ni and Zn in few stars could be attributed to a progenitor population being massive stars exploding as Type II supernovae \citep{nakamura1999, nomoto2013}. The uncertainties in the abundances of the program stars are also marked in figure 11.

The distribution of the neutron-capture elements Sr and Ba are shown in Figure 12. Strontium belongs to the first-peak or lighter n-capture elements, while Ba belongs to the heavier n-capture elements. Large enhancements or depletion in n-capture elements is not found among our program stars, with the exception of \sixfiftytwo. The ratio for light-to-heavy n-capture elements depends largely on the mass and nature of the progenitors. Since the contribution of the $s$-process is minimal at lower metallicities, the origin of these elements is expected to be from the $r$-process. Following \cite{tsuji1,tsuji2,susmitha} and \cite{siegel2019} for $r$-process origin, the heavier element Ba is produced primarily by neutron star mergers (NSMs) or collapsars, whereas the lighter element Sr can be synthesized in NSMs as well as Type II supernovae. Thus, an excess of one over the other indicates the dominance of either NSMs or core-collapse SNe, and thus provides valuable information about the nature of the progenitors for the origin of the $r$-process \citep{ban_rp}. For the present sample, the distribution of [Sr/Ba], as a function of [Fe/H], is shown in the bottom panel of Figure 12. From inspection, the scatter is very much lower, with a mean value of  $<Sr/Ba>$ = 0, making them likely to be polluted evenly from both the progenitor population during their star-forming epochs.

\begin{figure*}
\centering
\includegraphics[width=1.8\columnwidth]{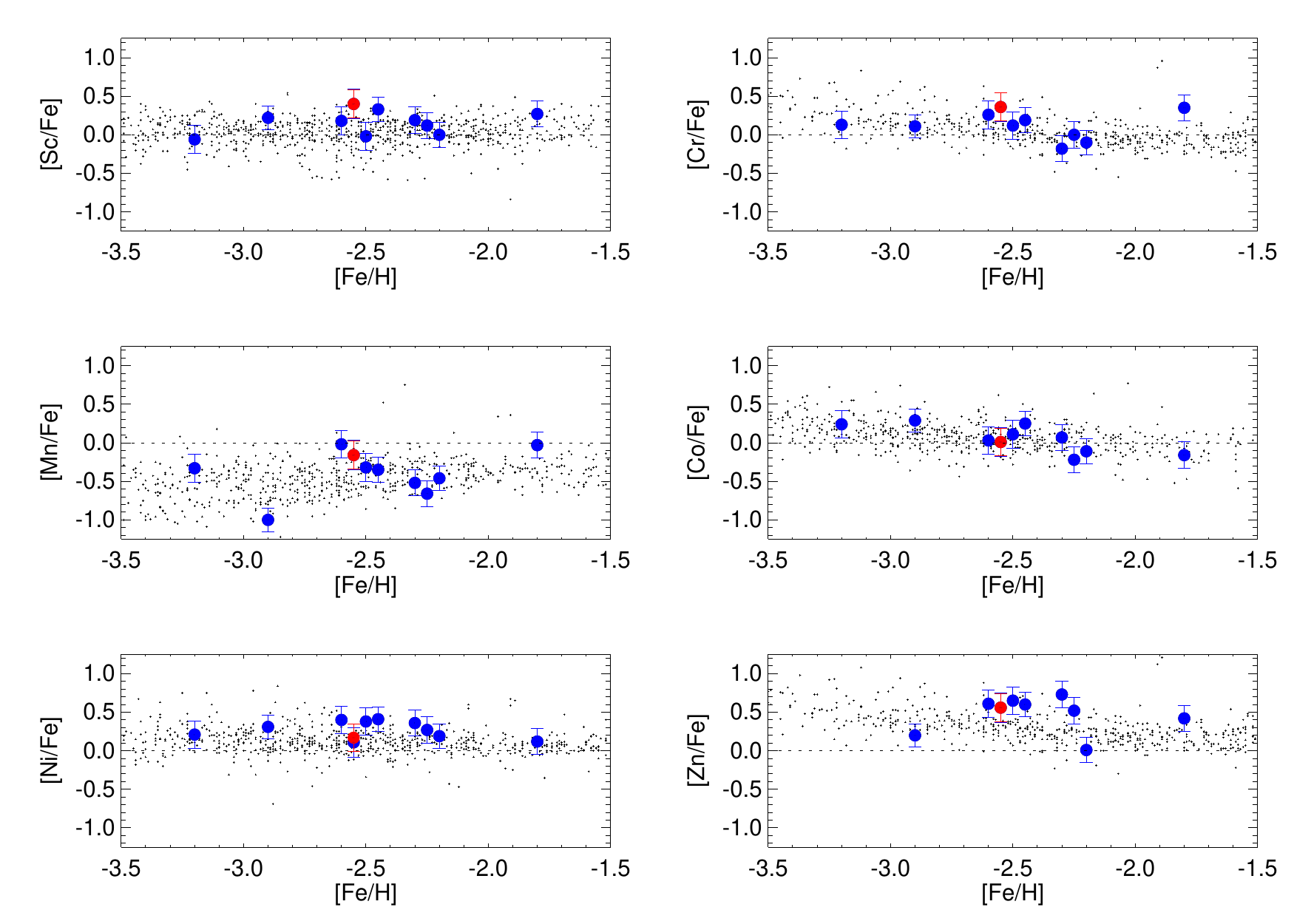}
\caption{The distribution of the Fe-peak elements Sc, Cr, Mn, Co, Ni, and Zn as a function of [Fe/H]. The blue-filled circles represent the program stars, while the red-filled circle denotes the abundances for \sixfiftytwo. The abundances of other metal-poor stars in literature compiled from \cite{sudasaga} are shown in black dots. The uncertainties are demonstrated by coloured error bars.}
\end{figure*}

\begin{figure}
\centering
\includegraphics[width=1.0\columnwidth]{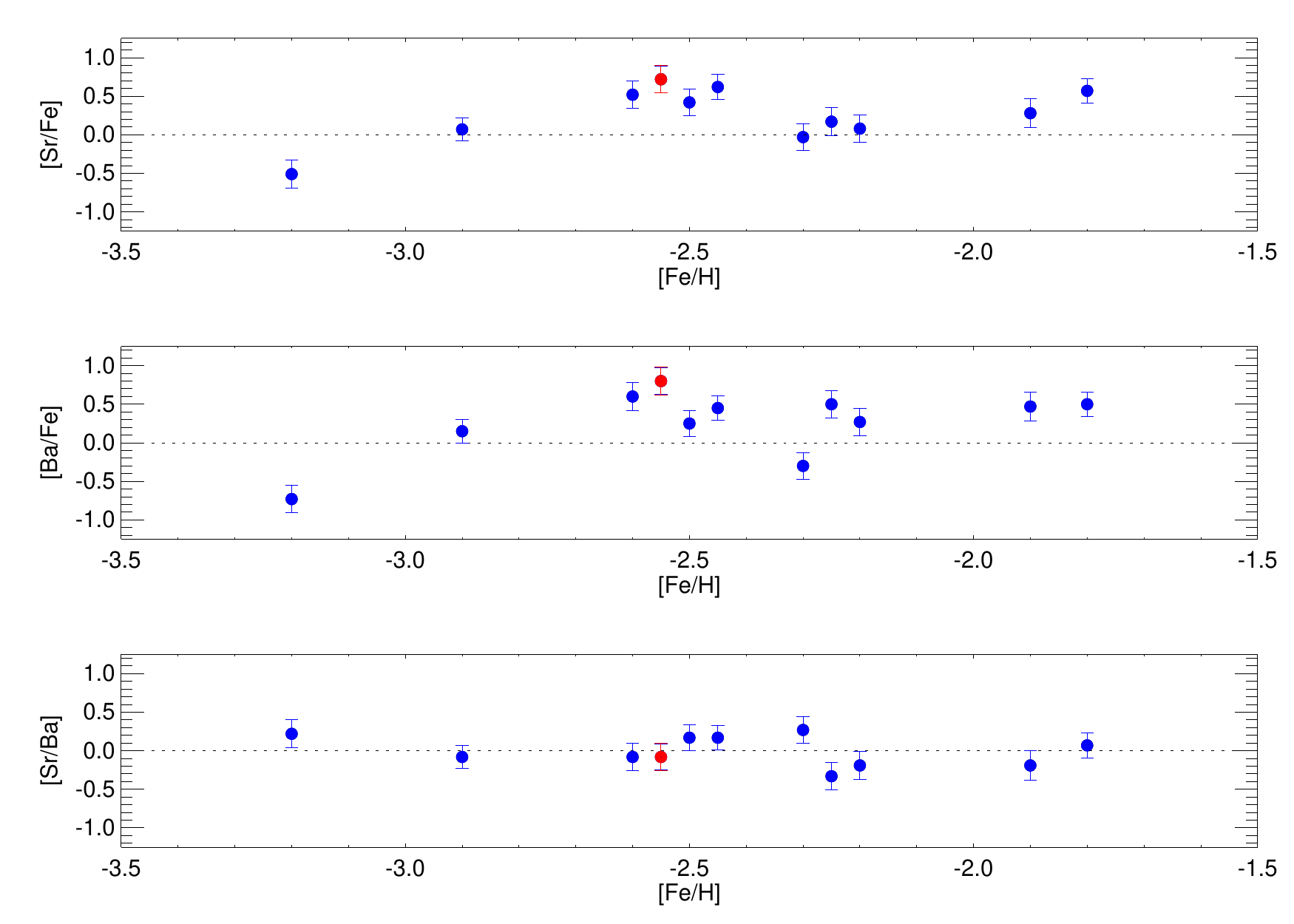}
\caption{The distribution of the neutron-capture elements Sr and Ba as a function of [Fe/H]. The blue-filled circles represent the program stars, while the red-filled circle denotes the abundances for \sixfiftytwo. The uncertainties are demonstrated by coloured error bars.}
\end{figure}


\subsection{Kinematics}

To study the assemblage history of the Milky Way, it is important to classify the origin of the stars based on their kinematics. The distribution of the Li-rich and Li-poor stars in the $L_z$ vs $L_{\perp}$ diagram can provide important clues towards the evolution of Li in the Milky Way. Following \cite{dimatteo2020}, stars with $L_z <$ –10 kpc km/s are dominated by those formed in situ while the stars in the region $L_{\perp} >$ 13 kpc km/s and $L_z >$ –10 kpc km/s are primarily accreted. However, the region $L_{\perp} <$ 13 kpc km/s and $L_z >$ –10 kpc km/s contains the stars from the Gaia-Sausage-Enceladus structure \citep{helmi2018,haywood2018}, as well as the disc stars with kinematics similar to the halo, and is therefore called the `mixed zone'. The three regions are shown by red dashed lines in Figure 13, while the previously discussed classification of prograde and retrograde motion stars is shown by the black dashed line. Four stars in the sample are found to have formed in situ, whereas one star is seen to be accreted. The rest of the sample belongs to the mixed zone. The position of the VMP and EMP stars in the Spite plateau from the SAGA database \cite{sudasaga} are shown by green circles in Figure 13. The filled circles indicate the stars in the Spite plateau with $A$(Li) $>$2.05, while the open green circles show the stars that are slightly depleted from the Spite plateau. The majority of the stars belong to the mixed zone, with very few populating either the in-situ or accreted zones. The Li-rich stars are primarily found towards the bottom of the mixed zone. Thus, the Li population in the Spite plateau has significant contributions from both stars formed in situ and those that are accreted. However, only two of the Li-rich stars in the Spite plateau are found in the accreted zone which is dominated by Li-depleted stars.

\begin{figure}
\centering
\includegraphics[width=1.0\columnwidth]{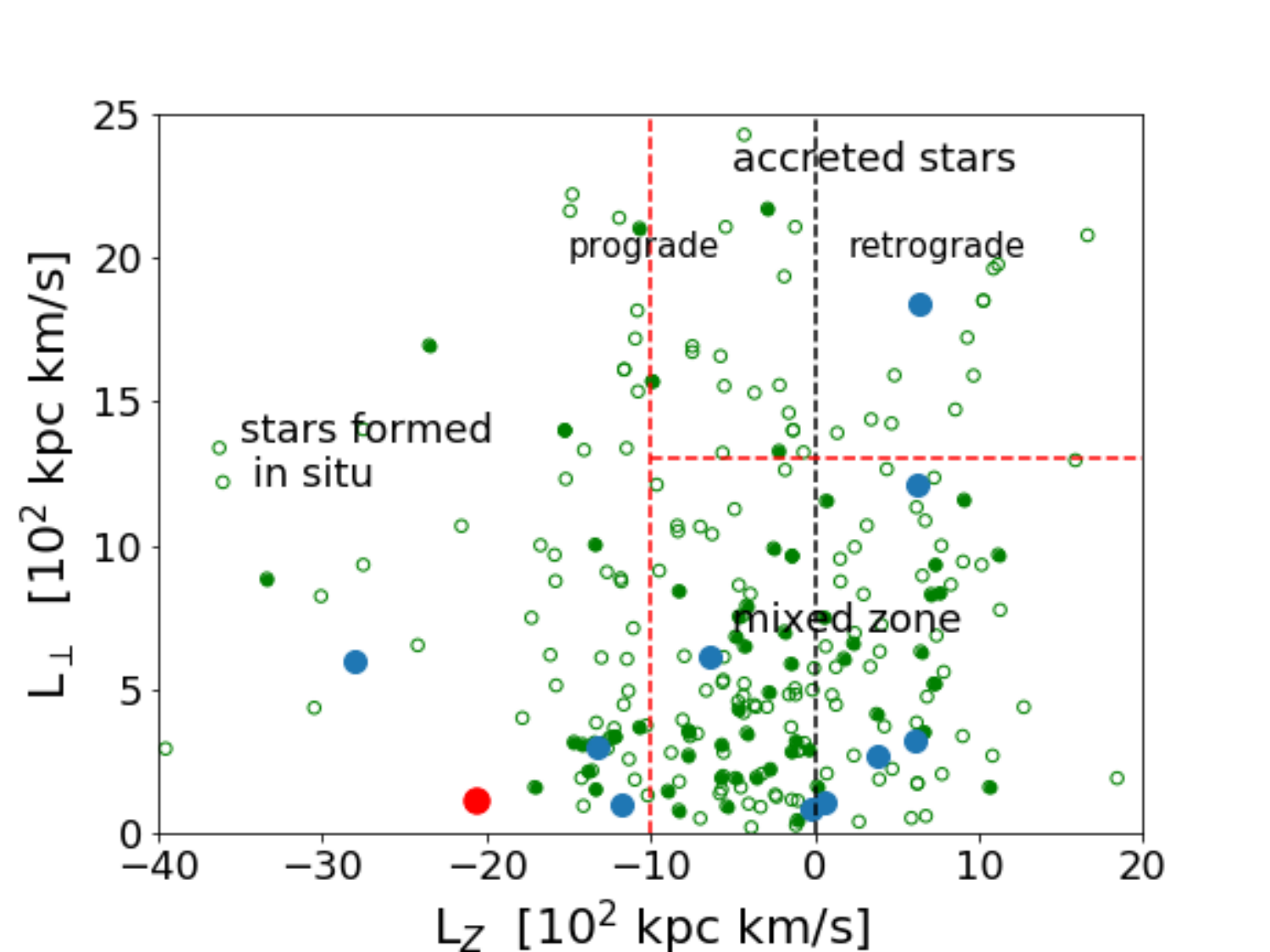}
\caption{Orbital characteristics of the program stars. The black dotted line separates prograde motions from retrograde motions, while the red dotted lines show the regions dominated by stars formed in situ, accreted stars, and in the mixed zone. The main sequence and turn-off VMP and EMP stars with Li measurement compiled from \cite{sudasaga} have been shown in green circles. The stars on the Spite plateau are shown as filled green circles, while slightly Li-poor stars are shown with green open circles.} 
\label{Lz_Lperp}
\end{figure}


\begin{figure*}
\centering
\includegraphics[width=1.8\columnwidth]{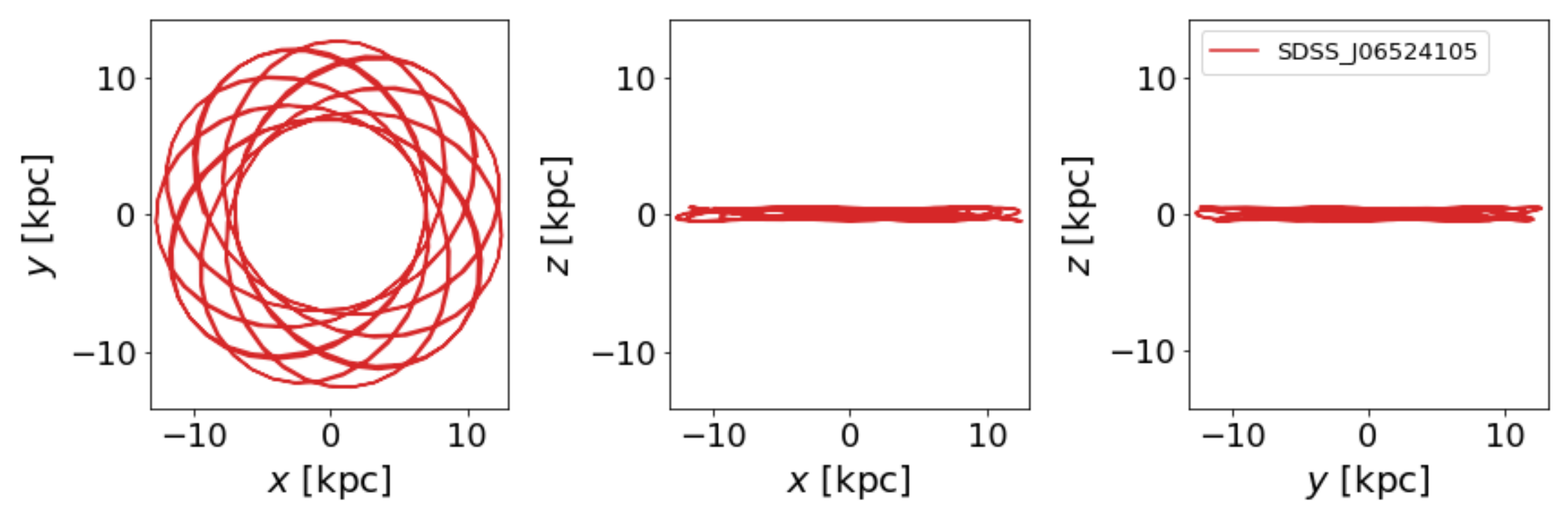}
\caption{The figure shows the trajectory of SDSS J0652+4105 in galactocentric coordinate which appears in the in-situ region of Fig. \ref{Lz_Lperp}. The orbit is shown in the XY, XZ and YZ planes. Trajectory has been calculated up to 5 Gyrs back in time from the present day.}
\label{insitu}
\end{figure*}

\begin{figure*}
\centering
\includegraphics[width=1.8\columnwidth]{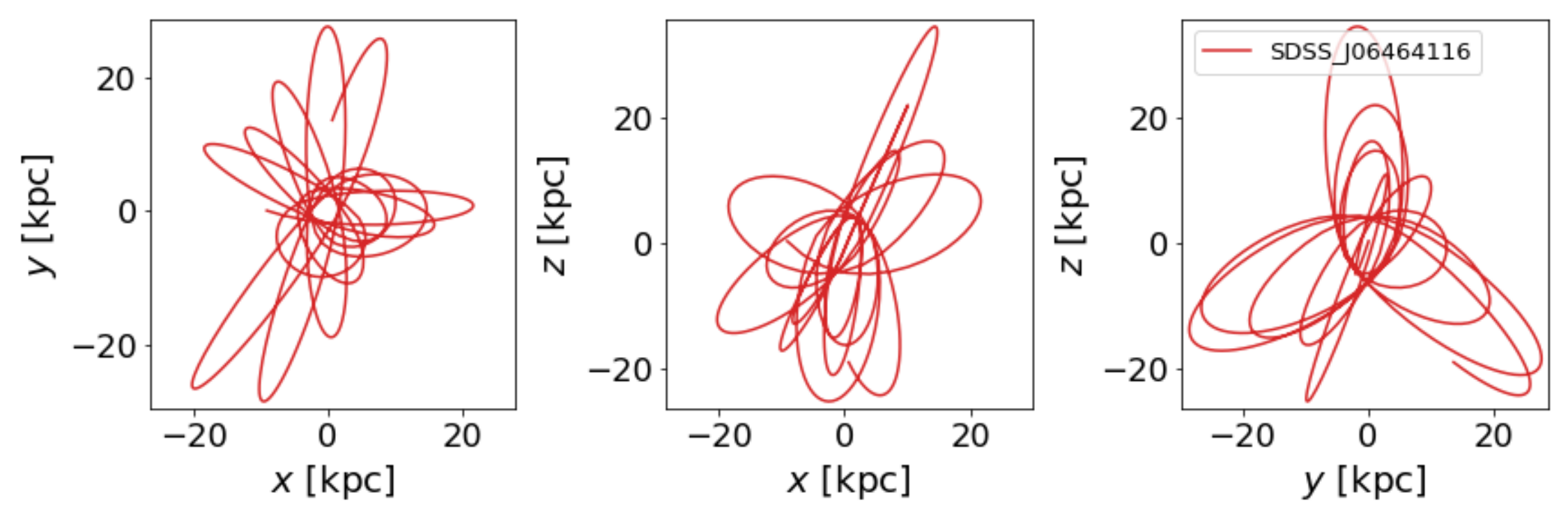}
\caption{The figure represents the same as Fig. \ref{insitu} for SDSS J0646+4116 which appears in the accreted region of Fig. \ref{Lz_Lperp}.}
\label{accreted}
\end{figure*}

\begin{figure*}
\centering
\includegraphics[width=1.34\columnwidth]{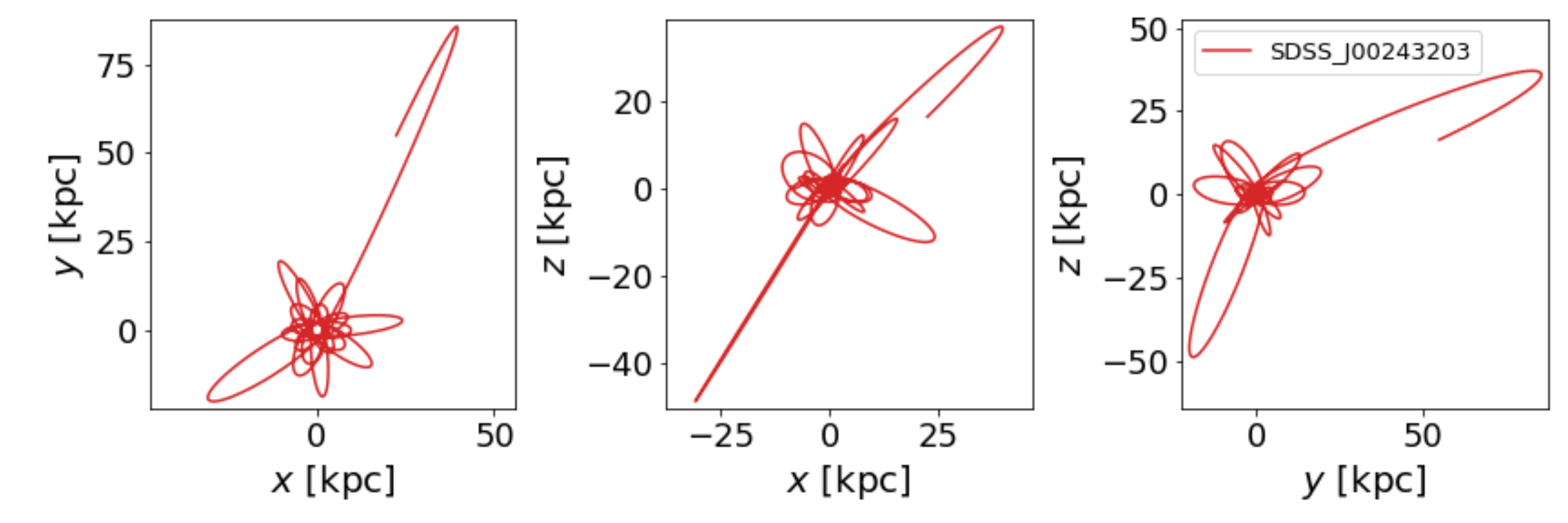}
\includegraphics[width=1.34\columnwidth]{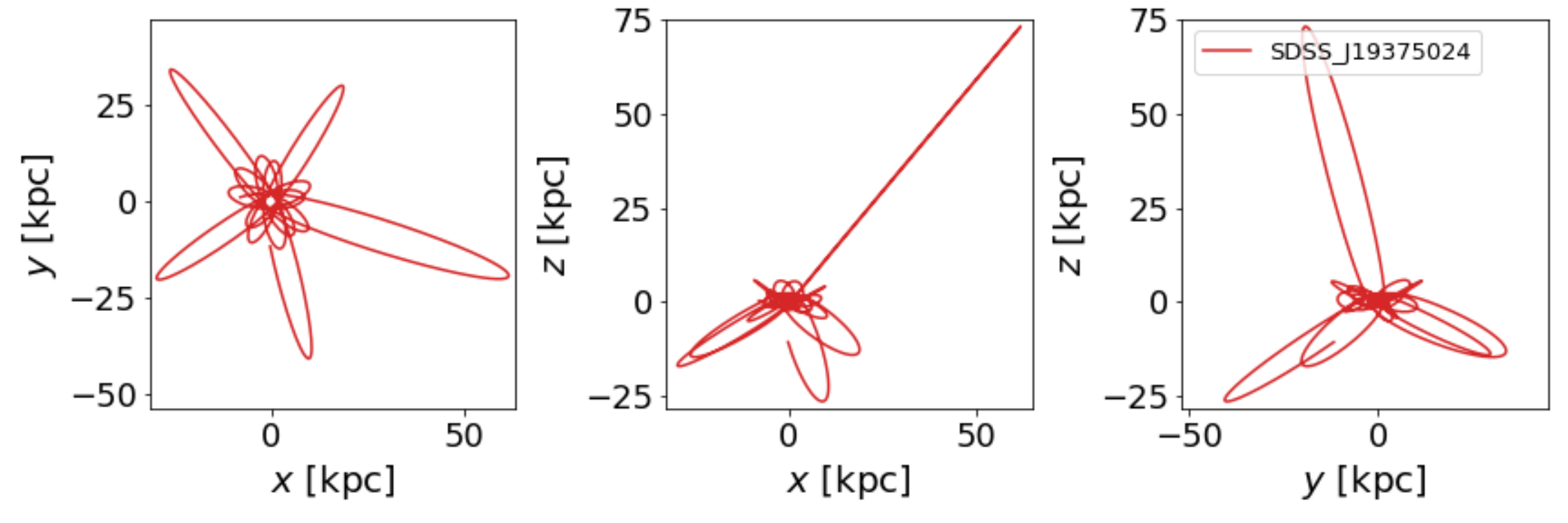}

\caption{ The figure represents the same as Fig. \ref{insitu} for two representative stars falling in the mixed zone of Fig. \ref{Lz_Lperp}. The names of the individual stars are mentioned on the respective figures.}
\label{mixed}
\end{figure*}

We have also calculated the trajectories for the stars in our sample for in-situ, accreted, and mixed zones in the classification as shown in Figure 13. The orbits are computed using $gala$ \footnote{The routine $gala$ is an Astropy affiliated package for Galactic dynamics. $gala$ provides functionality for representing analytic mass models that are commonly used in Galactic dynamics contexts for numerically integrating stellar orbits (e.g., Chapter 3 of Binney and Tremaine 2008). The gravitational potential models are defined by specifying parameters such as mass, scale radii, or shape parameters. Once defined, they can be used in combination with numerical integrators provided in $gala$ to compute orbits. $gala$ comes with a pre-defined, multi-component, but simple model for the Milky Way that can be used for orbit integrations.}. The trajectories were calculated for a time period of 5 Gyr. The orbits for the in-situ n-capture-rich star \sixfiftytwo is shown in Figure 14. The three panels indicate the motions in the XY, XZ, and YZ planes. The star is found to have almost no motion along the Z-axis, and thus is clearly formed in-situ as expected. Similarly, the orbit for the accreted star \zerosixgc is shown in Figure 15.  However, the trajectory appears very similar to that of stars formed in-situ, which could be possible, as \zerosixgc is also likely to be a globular cluster escapee. The mixed zone program stars in this figure appear to have open orbits and hence are most likely to be accreted rather than formed in-situ. The orbits for a few representative cases for the stars in the mixed zones are shown in Figure 16.

\section{Conclusion}

A sample of 9  stars in the domain of very metal-poor stars with weak molecular carbon CH $G$-bands, selected from the low-resolution SDSS/MARVELS pre-survey, have been observed at a high spectral resolution to study their detailed abundances. The stars show typical $\alpha$-element enhancements, and the odd-even nucleosynthesis pattern for the light elements. The Fe-peak elements mostly track the iron content; the observed trends are consistent with other metal-poor stars. Lithium could be detected and measured in all the program stars, and several belong to the Spite plateau. The depletion of Li is observed as the stars ascend the giant branch. The trends for depletion in Li with temperature are quantified; 85\% of the stars were found to fall within the 1$\sigma$ (0.19 and 0.12 dex/K for giants and dwarfs respectively) width of the best fit. A small slope of the Spite plateau at the metal-poor end was also found when the Li abundance was studied as a function of Galactocentric distance, indicating that the Li-rich MSTO stars are not preferentially located at larger distances from the Galactic plane. Most Li-rich stars are found to be in or close to the galactic plane. The stars have also been classified on the basis of their motion into prograde and retrograde samples. The program stars, along with the Spite plateau population in the literature, are divided into those that were likely formed in situ, accreted, and in the mixed zones. The orbits for the program stars have also been derived and studied for a period of 5 Gyr backwards in time. The mixed zone is found to be the most populated, and thus neither formation in situ nor by accretion, and can be considered as an important contributor to the population of the Spite plateau. 

\section{Acknowledgements}

We thank the staff of IAO, Hanle, and CREST, Hosakote, that made these observations possible. The facilities at IAO and CREST are operated by the Indian Institute of Astrophysics, Bangalore. T.C.B. acknowledges partial support from grant PHY 14-30152 (Physics Frontier Center/JINA-CEE), awarded by the U.S. National Science Foundation (NSF). We also thank the anonymous referee for the comments which improved the quality of our paper.


\begin{table}
\tabcolsep3.0pt $ $
\begin{center}
\caption{Elemental-abundance determinations for SDSS J0024+3203} 
\begin{tabular}{cccccccc}
\hline\hline
Elements &Species & $N_{lines}$ & A(X) & Solar & [X/H] & [X/Fe] & $\sigma$ \\
\hline

Li &Li I &1 &2.00 \\
C &CH & \dots    &6.00 &8.43   &$-$2.43 &0.02 &synth \\
Na$^b$ &Na I &2 &4.22 &6.21        &$-$1.99 &0.46 &synth \\
Mg &Mg I &4 &5.64 &7.59        &$-$1.95 &0.50 &synth\\
Al$^b$ &Al I &1 &3.90 &6.43        &$-$2.53 &$-$0.08 &synth \\
Si &Si I &2 &5.46 &7.51 &$-$2.05 &0.40 &0.09 \\
Ca &Ca I &8 &4.37 &6.32        &$-$1.95 &0.50 &0.06\\
Sc &Sc II &5 &1.03 &3.15    &$-$2.12 &0.33 &0.06\\
Ti &Ti I &7 &3.14 &4.93        &$-$1.79 &0.66 &0.09\\
   &Ti II &6 &3.03 &4.93       &$-$1.90 &0.55 &0.05\\
Cr &Cr I &3 &3.21 &5.62        &$-$2.41 &0.04 &0.12\\
   &Cr II &2 &3.52 &5.62       &$-$2.10 &0.35 &0.07\\
Mn &Mn I &4 &2.62 &5.42        &$-$2.80 &$-$0.35 &0.11\\
Co &Co I &2 &2.73 &4.93        &$-$2.20 &0.25 &0.06\\
Ni &Ni I &3 &4.16 &6.20        &$-$2.04 &0.41 &synth\\
Zn &Zn I &2 &2.71 &4.56        &$-$1.85 &0.60 &0.07\\
Sr &Sr II &2 &1.00 &2.83    &$-$1.83 &0.62 &synth \\
Ba &Ba II &2 &0.25 &2.25    &$-$2.00 &0.45 &synth \\
Eu$^u$ &Eu II &1 &$-$1.0 &0.52 &$-$1.52 &0.93 &synth \\

\hline
\end{tabular}
\end{center}
    $\sigma$ indicates the random error.
         \newline
    $^b$ Values obtained after applying NLTE corrections.
        \newline
    $^u$ indicates an upper limit.
\end{table}
 
 \begin{table}
 \tabcolsep3.0pt $ $
\begin{center}
\caption{Elemental-abundance determinations for SDSS J0315+2123} \label{c6t4}
\begin{tabular}{crrrrrrr}
\hline\hline
Elements &Species & $N_{lines}$ & A(X) & Solar & [X/H] & [X/Fe] & $\sigma$ \\
\hline

Li &Li I &1 &1.80 \\
C &CH & \dots    &6.10 &8.43   &$-$2.33 &$-$0.03 &synth \\
Na$^b$ &Na I &2 &3.74 &6.21        &$-$2.47 &$-$0.17 &synth \\
Mg &Mg I &4 &5.85 &7.59        &$-$1.74 &0.56 &synth\\
Si &Si I &1 &5.77 &7.51        &$-$1.74 &0.56 &synth \\
Ca &Ca I &8 &4.24 &6.34        &$-$2.10 &0.20 &0.08\\
Sc &Sc II &5 &$-$1.04 &3.15    &$-$2.11 &0.19 &0.04\\
Ti &Ti I &7 &3.11 &4.93        &$-$1.82 &0.48 &0.09\\
   &Ti II &6 &2.98 &4.93       &$-$1.95 &0.35 &0.13\\
Cr &Cr I &3 &3.08 &5.62        &$-$2.54 &$-$0.22 &0.08\\
   &Cr II &2 &3.19 &5.62       &$-$2.43 &$-$0.13 &0.09\\
Mn &Mn I &4 &2.60 &5.42        &$-$2.82 &$-$0.52 &0.12\\
Co &Co I &2 &2.70 &4.93        &$-$2.23 &0.07 &0.06\\
Ni &Ni I &3 &4.26 &6.20        &$-$1.94 &0.36 &synth\\
Cu &Cu I &2 &2.32 &4.19        &$-$1.87 &0.43 &0.12 \\
Zn &Zn I &2 &3.09 &4.56        &$-$1.47 &0.83 &synth\\
Sr &Sr II &2 &0.40 &2.83    &$-$2.43 &$-$0.03 &synth \\
Ba &Ba II &2 &0.25 &2.25    &$-$2.00 &$-$0.30 &synth \\

\hline
\end{tabular}
\end{center}
\end{table}
 
 \begin{table}
 \tabcolsep3.0pt $ $
\begin{center}
\caption{Elemental-abundance determinations for SDSS J0643+5934} \label{c6t9}
\begin{tabular}{crrrrrrr}
\hline\hline
Elements &Species & $N_{lines}$ & A(X) & Solar & [X/H] & [X/Fe] & $\sigma$ \\
\hline

Li &Li I &1 &0.80 \\
C &CH & \dots    &5.75 &8.43   &$-$2.68 &0.22 &synth \\
Na$^b$ &Na I &2 &3.54 &6.21        &$-$2.67 &0.23 &synth \\
Mg &Mg I &4 &5.03 &7.59        &$-$2.56 &0.34 &synth\\
Al$^b$ &Al I &1 &2.70 &6.43        &$-$3.73 &$-$0.83 &0.09 \\
Ca &Ca I &8 &3.61 &6.32        &$-$2.71 &0.19 &0.06\\
Sc &Sc II &5 &0.47 &3.15    &$-$2.68 &0.22 &0.11\\
Ti &Ti I &7 &2.42 &4.93        &$-$2.51 &0.39 &0.10\\
   &Ti II &6 &2.36 &4.93       &$-$2.57 &0.33 &0.07\\
Cr &Cr I &3 &2.54 &5.62        &$-$3.08 &$-$0.18 &0.08\\
   &Cr II &2 &3.02 &5.62       &$-$2.60 &0.30 &0.08\\
Mn &Mn I &4 &1.52 &5.42        &$-$3.90 &$-$1.00 &0.12\\
Co &Co I &2 &2.32 &4.93        &$-$2.61 &0.29 &0.11\\
Ni &Ni I &3 &3.61 &6.20        &$-$2.59 &0.31 &synth\\
Zn &Zn I &2 &1.86 &4.56        &$-$2.70 &0.20 &0.06\\
Sr &Sr II &2 &0.00 &2.83    &$-$2.83 &0.07 &synth \\
Ba &Ba II &2 &$-$0.50 &2.25    &$-$2.75 &0.15 &synth \\

\hline
\end{tabular}
\end{center}
\end{table}

\begin{table}
\tabcolsep3.0pt $ $
\begin{center}
\caption{Elemental-abundance determinations for SDSS J0652+4105} \label{c5t5}
\begin{tabular}{ccrrrrrr}
\hline\hline
Elements &Species & $N_{lines}$ & A(X) & Solar & [X/H] & [X/Fe] & $\sigma$ \\
\hline
Li &Li I &1 &1.75 & & & &synth \\
C &CH &    &5.75 &8.43 &$-$2.68 &$-$0.13 &synth \\
Na$^b$ &Na I &2 &4.23 &6.21 &$-$1.98 &0.57 &synth \\
Mg &Mg I &5 &5.62 &7.59 &$-$1.97 &0.58 &synth \\
Al$^b$ &Al I &1 &2.96 &6.43 &$-$3.47 &$-$0.92 &0.18\\
Ca &Ca I &11 &4.14 &6.32 &$-$2.18 &0.37 &0.08\\
Sc &Sc II &5 &1.00 &3.15 &$-$2.15 &0.40 &0.10\\
Ti &Ti I &4 &2.94 &4.93 &$-$1.99 &0.56 &0.15\\
   &Ti II &13 &2.78 &4.93 &$-$2.15 &0.40 &0.11\\
Cr &Cr I &6 &3.24 &5.62 &$-$2.38 &0.17 &0.16\\
   &Cr II &1 &3.69 &5.62 &$-$1.93 &0.62 &0.09\\
Mn &Mn I &5 &2.71 &5.42 &$-$2.71 &$-$0.16 &0.12\\
Co &Co I &2 &2.39 &4.93 &$-$2.54 &0.01 &0.08\\
Ni &Ni I &4 &3.82 &6.20 &$-$2.38 &0.17 &synth \\
Zn &Zn I &1 &2.57 &4.56 &$-$1.99 &0.56 &synth \\
Sr &Sr II &2 &1.00 &2.83 &$-$1.83 &0.72 &synth\\
Y &Y II &2 &0.25 &2.21 &$-$1.96 &0.59 &synth \\
Zr &Zr II &3 &0.75 &2.59 &$-$1.84 &0.71 &synth \\
Ba &Ba II &2 &0.50 &2.25 &$-$1.75 &0.80 &synth \\
La &La II &2 &$-$0.87 &1.11 &$-$1.98 &0.57 &synth\\
Nd &Nd II &2 &0.0 &1.42 &$-$1.42 &1.13 &synth \\
Eu &Eu II &1 &$-$1.0 &0.52 &$-$1.52 &1.03 &synth \\
\hline
\end{tabular}
\end{center}

\end{table}

\begin{table}
\tabcolsep3.0pt $ $
\begin{center}
\caption{Elemental-abundance determinations for SDSS J1024+4151} \label{c6t12}
\begin{tabular}{crrrrrrrrrrr}
\hline\hline
Elements &Species & $N_{lines}$ & A(X) & Solar & [X/H] & [X/Fe] & $\sigma$ \\
\hline

Li &Li I &1 &1.05 \\
C &CH & \dots    &6.00 &8.43   &$-$2.43 &$-$0.18 &synth \\
O &O I & \dots &8.00 &8.69      &$-$00.69 &1.56 &synth \\
Na$^b$ &Na I &2 &3.90 &6.21        &$-$2.31 &$-$0.06 &synth \\
Mg &Mg I &4 &5.76 &7.59        &$-$2.03 &0.22 &synth\\
Al$^b$ &Al I &1 &2.89 &6.43        &$-$3.54 &$-$1.29 &synth \\
Ca &Ca I &8 &3.68 &6.32        &$-$2.64 &0.46 &0.09\\
Si &Si I &2 &5.42 &7.51        &$-$2.09 &0.14 &0.16 \\
Sc &Sc II &5 &1.08 &3.15       &$-$2.23 &0.12 &0.08\\
Ti &Ti I &7 &3.42 &4.93        &$-$1.51 &0.74 &0.12\\
   &Ti II &6 &3.11 &4.93       &$-$1.82 &0.43 &0.09\\
Cr &Cr I &3 &3.25 &5.62        &$-$2.37 &$-$0.12 &0.13\\
   &Cr II &2 &3.50 &5.62       &$-$2.12 &0.13 &0.09\\
Mn &Mn I &4 &2.51 &5.42        &$-$2.91 &$-$0.66 &0.10\\
Co &Co I &2 &2.46 &4.93        &$-$2.47 &$-$0.22 &0.06\\
Ni &Ni I &3 &4.22 &6.20        &$-$1.98 &0.27 &synth\\
Cu &Cu I &1 &2.89 &4.56 &$-$1.67 &0.58 &synth\\
Zn &Zn I &2 &2.89 &4.56        &$-$1.67 &0.58 &0.11\\
Sr &Sr II &2 &0.75 &2.83    &$-$2.08 &0.17 &synth \\
Ba &Ba II &2 &0.50 &2.25    &$-$1.75 &0.50 &synth \\
Eu$^u$ &Eu II &1 &$-$0.75 &0.52 &$-$1.27 &0.98 &synth \\

\hline
\end{tabular}
\end{center}
\end{table}

\begin{table}
\tabcolsep3.0pt $ $
\begin{center}
\caption{Elemental-abundance determinations for SDSS J1146+2343} \label{c6t2}
\begin{tabular}{crrrrrrr}
\hline\hline
Elements &Species & $N_{lines}$ & A(X) & Solar & [X/H] & [X/Fe] & $\sigma$ \\
\hline
Li &Li I &1 &1.15 \\
C &CH & \dots    &6.00 &8.43   &$-$2.43 &$-$0.17 &synth \\
Na$^b$ &Na I &2 &3.67 &6.21        &$-$2.64 &$-$0.04 &synth \\
Mg &Mg I &4 &5.39 &7.59        &$-$2.00 &0.60 &synth\\
Al$^b$ &Al I &1 &2.90 &6.43        &$-$3.53 &$-$0.93 &synth \\
Ca &Ca I &8 &4.10 &6.32        &$-$2.22 &0.38 &0.06\\
Sc &Sc II &5 &$-$0.73 &3.15    &$-$2.42 &0.18 &0.01\\
Ti &Ti I &7 &2.83 &4.93        &$-$2.10 &0.50 &0.03\\
   &Ti II &6 &2.62 &4.93       &$-$2.31 &0.29 &0.04\\
Cr &Cr I &3 &3.07 &5.62        &$-$2.55 &0.05 &0.05\\
   &Cr II &2 &3.49 &5.62       &$-$2.13 &0.47 &0.05\\
Mn &Mn I &4 &2.84 &5.42        &$-$2.58 &$-$0.02 &0.02\\
Co &Co I &2 &2.36 &4.93        &$-$2.57 &0.03 &0.01\\
Ni &Ni I &3 &4.00 &6.20        &$-$2.20 &0.40 &synth\\
Zn &Zn I &2 &2.57 &4.56        &$-$1.99 &0.61 &0.05\\
Sr &Sr II &2 &0.75 &2.83    &$-$2.08 &0.52 &synth \\
Ba &Ba II &2 &0.25 &2.25    &$-$2.00 &0.60 &synth \\
Eu$^u$ &Eu II &1 &$-$1.25 &0.52 &$-$1.77 &0.83 &synth \\

\hline
\end{tabular}
\end{center}
    $\sigma$ indicates the random error.
    $^b$ Values obtained after applying NLTE corrections.
    $^u$ indicates an upper limit.
\end{table}

\begin{table}
\tabcolsep3.0pt $ $
\begin{center}
\caption{Elemental-abundance determinations for SDSS J1725+4202} 
\begin{tabular}{crrrrrrr}
\hline\hline
Elements &Species & $N_{lines}$ & A(X) & Solar & [X/H] & [X/Fe] & $\sigma$ \\
\hline

Li &Li I &1 &1.90 \\
C &CH & \dots    &6.00 &8.43   &$-$2.43 &0.07 &synth \\
Na$^b$ &Na I &2 &3.92 &6.21        &$-$2.29 &0.21 &synth \\
Mg &Mg I &4 &5.53 &7.59        &$-$2.06 &0.44 &synth\\
Al$^b$ &Al I &1 &3.37 &6.43        &$-$3.06 &$-$0.56 &synth \\
Ca &Ca I &8 &4.28 &6.32        &$-$2.02 &0.48 &0.08\\
Sc &Sc II &5 &0.63 &3.15    &$-$2.52 &$-$0.02 &0.09\\
Ti &Ti I &7 &2.88 &4.93        &$-$2.05 &0.45 &0.08\\
   &Ti II &6 &2.87 &4.93       &$-$2.06 &0.44 &0.04\\
Cr &Cr I &3 &3.03 &5.62        &$-$2.59 &$-$0.09 &0.07\\
   &Cr II &2 &3.46 &5.62       &$-$2.16 &0.34 &0.11\\
Mn &Mn I &4 &2.60 &5.42        &$-$2.92 &$-$0.32 &0.15\\
Co &Co I &2 &2.54 &4.93        &$-$2.39 &0.11 &0.09\\
Ni &Ni I &3 &4.08 &6.20        &$-$2.12 &0.38 &synth\\
Zn &Zn I &2 &2.71 &4.56        &$-$1.85 &0.65 &synth\\
Sr &Sr II &2 &0.75 &2.83    &$-$2.08 &0.42 &synth \\
Ba &Ba II &2 &0.00 &2.25    &$-$2.25 &0.25 &synth \\

\hline
\end{tabular}
\end{center}
\end{table}

\begin{table}
\tabcolsep3.0pt $ $
\begin{center}
\caption{Elemental-abundance determinations for SDSS J1933+4524} \label{c6t8}
\begin{tabular}{cccccrrr}
\hline\hline
Elements &Species & $N_{lines}$ & A(X) & Solar & [X/H] & [X/Fe] & $\sigma$ \\
\hline

Li &Li I &1 &2.25 \\
C &CH & \dots    &6.50 &8.43   &$-$1.93 &$-$0.13 &synth \\
Na$^b$ &Na I &2 &4.23 &6.21    &$-$1.98 &$-$0.18 &synth \\
Mg &Mg I &4 &5.95 &7.59        &$-$1.64 &0.16 &synth\\
Al$^b$ &Al I &1 &3.85 &6.43    &$-$2.58 &$-$0.78 &synth \\
Ca &Ca I &8 &4.86 &6.32        &$-$1.46 &0.34 &0.08\\
Sc &Sc II &5 &1.62 &3.15       &$-$1.53 &0.27 &0.09\\
Ti &Ti I &7 &3.60 &4.93        &$-$1.33 &0.47 &0.08\\
   &Ti II &6 &3.75 &4.93       &$-$1.18 &0.62 &0.04\\
Cr &Cr I &3 &4.18 &5.62        &$-$1.44 &0.36 &0.07\\
   &Cr II &2 &4.15 &5.62       &$-$1.47 &0.33 &0.11\\
Mn &Mn I &4 &3.59 &5.42        &$-$1.83 &$-$0.03 &0.15\\
Co &Co I &2 &2.97 &4.93        &$-$1.96 &$-$0.16 &0.09\\
Ni &Ni I &3 &4.52 &6.20        &$-$1.68 &0.12 &synth\\
Zn &Zn I &2 &3.18 &4.56        &$-$1.38 &0.42 &synth\\
Sr &Sr II &2 &1.60 &2.83       &$-$1.23 &0.57 &synth \\
Ba &Ba II &2 &0.95 &2.25       &$-$1.30 &0.50 &synth \\
Eu$^u$ &Eu II &1 &$-$0.50 &0.52 &$-$1.02 &0.78 &synth \\

\hline
\end{tabular}
\end{center}
\end{table}

\bibliographystyle{aasjournal}
\bibliography{ms_li_apj}

\begin{thebibliography}{}
\expandafter\ifx\csname natexlab\endcsname\relax\def\natexlab#1{#1}\fi
\providecommand{\url}[1]{\href{#1}{#1}}
\providecommand{\dodoi}[1]{doi:~\href{http://doi.org/#1}{\nolinkurl{#1}}}
\providecommand{\doeprint}[1]{\href{http://ascl.net/#1}{\nolinkurl{http://ascl.net/#1}}}
\providecommand{\doarXiv}[1]{\href{https://arxiv.org/abs/#1}{\nolinkurl{https://arxiv.org/abs/#1}}}

\bibitem[{{Alvarez} \& {Plez}(1998)}]{alvarezplez1998}
{Alvarez}, R., \& {Plez}, B. 1998, \aap, 330, 1109

\bibitem[{{Andrievsky} {et~al.}(2007){Andrievsky}, {Spite}, {Korotin}, {Spite},
  {Bonifacio}, {Cayrel}, {Hill}, \& {Fran{\c{c}}ois}}]{andrievskyna}
{Andrievsky}, S.~M., {Spite}, M., {Korotin}, S.~A., {et~al.} 2007, \aap, 464,
  1081, \dodoi{10.1051/0004-6361:20066232}

\bibitem[{{Andrievsky} {et~al.}(2008){Andrievsky}, {Spite}, {Korotin}, {Spite},
  {Bonifacio}, {Cayrel}, {Hill}, \& {Fran{\c{c}}ois}}]{andrievskyal}
---. 2008, \aap, 481, 481, \dodoi{10.1051/0004-6361:20078837}

\bibitem[{{Aoki} {et~al.}(2009){Aoki}, {Barklem}, {Beers}, {Christlieb},
  {Inoue}, {Garc{\'{\i}}a P{\'e}rez}, {Norris}, \& {Carollo}}]{aoki2009}
{Aoki}, W., {Barklem}, P.~S., {Beers}, T.~C., {et~al.} 2009, \apj, 698, 1803,
  \dodoi{10.1088/0004-637X/698/2/1803}

\bibitem[{{Bailer-Jones} {et~al.}(2021){Bailer-Jones}, {Rybizki}, {Fouesneau},
  {Demleitner}, \& {Andrae}}]{bailerjones2021}
{Bailer-Jones}, C.~A.~L., {Rybizki}, J., {Fouesneau}, M., {Demleitner}, M., \&
  {Andrae}, R. 2021, \aj, 161, 147, \dodoi{10.3847/1538-3881/abd806}

\bibitem[{Bandyopadhyay {et~al.}(2020{\natexlab{a}})Bandyopadhyay, Sivarani, \&
  Beers}]{ban_rp}
Bandyopadhyay, A., Sivarani, T., \& Beers, T.~C. 2020{\natexlab{a}}, The
  Astrophysical Journal, 899, 22, \dodoi{10.3847/1538-4357/ab9c9d}

\bibitem[{{Bandyopadhyay} {et~al.}(2018){Bandyopadhyay}, {Sivarani},
  {Susmitha}, {Beers}, {Giridhar}, {Surya}, \& {Masseron}}]{bandyopadhyay}
{Bandyopadhyay}, A., {Sivarani}, T., {Susmitha}, A., {et~al.} 2018, \apj, 859,
  114, \dodoi{10.3847/1538-4357/aabe80}

\bibitem[{Bandyopadhyay {et~al.}(2020{\natexlab{b}})Bandyopadhyay, Thirupathi,
  Beers, \& Susmitha}]{ban_gce}
Bandyopadhyay, A., Thirupathi, S., Beers, T.~C., \& Susmitha, A.
  2020{\natexlab{b}}, Monthly Notices of the Royal Astronomical Society, 494,
  36–43, \dodoi{10.1093/mnras/staa671}

\bibitem[{{Bayo} {et~al.}(2008){Bayo}, {Rodrigo}, {Barrado Y Navascu{\'e}s},
  {Solano}, {Guti{\'e}rrez}, {Morales-Calder{\'o}n}, \&
  {Allard}}]{bayo2008vosa}
{Bayo}, A., {Rodrigo}, C., {Barrado Y Navascu{\'e}s}, D., {et~al.} 2008, \aap,
  492, 277, \dodoi{10.1051/0004-6361:200810395}

\bibitem[{{Beers} \& {Christlieb}(2005)}]{beers2005}
{Beers}, T.~C., \& {Christlieb}, N. 2005, \araa, 43, 531,
  \dodoi{10.1146/annurev.astro.42.053102.134057}

\bibitem[{Bergemann \& Cescutti(2010)}]{bergemanncescutti2010}
Bergemann, M., \& Cescutti, G. 2010, Astronomy \& Astrophysics, 522, A9,
  \dodoi{10.1051/0004-6361/201014250}

\bibitem[{{Bergemann} \& {Gehren}(2008)}]{bergemann2008}
{Bergemann}, M., \& {Gehren}, T. 2008, \aap, 492, 823,
  \dodoi{10.1051/0004-6361:200810098}

\bibitem[{{Bergemann, Maria} {et~al.}(2019){Bergemann, Maria}, {Gallagher,
  Andrew J.}, {Eitner, Philipp}, {Bautista, Manuel}, {Collet, Remo},
  {Yakovleva, Svetlana A.}, {Mayriedl, Anja}, {Plez, Bertrand}, {Carlsson,
  Mats}, {Leenaarts, Jorrit}, {Belyaev, Andrey K.}, \& {Hansen,
  Camilla}}]{bergemann2019}
{Bergemann, Maria}, {Gallagher, Andrew J.}, {Eitner, Philipp}, {et~al.} 2019,
  A\&A, 631, A80, \dodoi{10.1051/0004-6361/201935811}

\bibitem[{{Bonifacio} {et~al.}(2007{\natexlab{a}}){Bonifacio}, {Molaro},
  {Sivarani}, {Cayrel}, {Spite}, {Spite}, {Plez}, {Andersen}, {Barbuy},
  {Beers}, {Depagne}, {Hill}, {Fran{\c c}ois}, {Nordstr{\"o}m}, \&
  {Primas}}]{firststars7}
{Bonifacio}, P., {Molaro}, P., {Sivarani}, T., {et~al.} 2007{\natexlab{a}},
  \aap, 462, 851, \dodoi{10.1051/0004-6361:20064834}

\bibitem[{{Bonifacio} {et~al.}(2007{\natexlab{b}}){Bonifacio}, {Molaro},
  {Sivarani}, {Cayrel}, {Spite}, {Spite}, {Plez}, {Andersen}, {Barbuy},
  {Beers}, {Depagne}, {Hill}, {Fran{\c c}ois}, {Nordstr{\"o}m}, \&
  {Primas}}]{bonifacio2007}
---. 2007{\natexlab{b}}, \aap, 462, 851, \dodoi{10.1051/0004-6361:20064834}

\bibitem[{{Bonifacio} {et~al.}(2009){Bonifacio}, {Spite}, {Cayrel}, {Hill},
  {Spite}, {Fran{\c{c}}ois}, {Plez}, {Ludwig}, {Caffau}, {Molaro}, {Depagne},
  {Andersen}, {Barbuy}, {Beers}, {Nordstr{\"o}m}, \& {Primas}}]{bonifacio2009}
{Bonifacio}, P., {Spite}, M., {Cayrel}, R., {et~al.} 2009, \aap, 501, 519,
  \dodoi{10.1051/0004-6361/200810610}

\bibitem[{{Bonifacio} {et~al.}(2015){Bonifacio}, {Caffau}, {Spite}, {Limongi},
  {Chieffi}, {Klessen}, {Fran{\c c}ois}, {Molaro}, {Ludwig}, {Zaggia}, {Spite},
  {Plez}, {Cayrel}, {Christlieb}, {Clark}, {Glover}, {Hammer}, {Koch},
  {Monaco}, {Sbordone}, \& {Steffen}}]{bonifacio2015}
{Bonifacio}, P., {Caffau}, E., {Spite}, M., {et~al.} 2015, \aap, 579, A28,
  \dodoi{10.1051/0004-6361/201425266}

\bibitem[{Brown {et~al.}(2021)Brown, Vallenari, Prusti, de~Bruijne, Babusiaux,
  Biermann, Creevey, Evans, Eyer, \& et~al.}]{gaia_edr3}
Brown, A. G.~A., Vallenari, A., Prusti, T., {et~al.} 2021, Astronomy \&
  Astrophysics, 650, C3, \dodoi{10.1051/0004-6361/202039657e}

\bibitem[{{Cameron} \& {Fowler}(1971)}]{cameronfowler1971}
{Cameron}, A.~G.~W., \& {Fowler}, W.~A. 1971, \apj, 164, 111,
  \dodoi{10.1086/150821}

\bibitem[{{Castelli} \& {Kurucz}(2004)}]{castellikurucz}
{Castelli}, F., \& {Kurucz}, R.~L. 2004, ArXiv Astrophysics e-prints

\bibitem[{{Cayrel} {et~al.}(2004){Cayrel}, {Depagne}, {Spite}, {Hill}, {Spite},
  {Fran{\c c}ois}, {Plez}, {Beers}, {Primas}, {Andersen}, {Barbuy},
  {Bonifacio}, {Molaro}, \& {Nordstr{\"o}m}}]{cayrel2004}
{Cayrel}, R., {Depagne}, E., {Spite}, M., {et~al.} 2004, \aap, 416, 1117,
  \dodoi{10.1051/0004-6361:20034074}

\bibitem[{{Cayrel de Strobel} \& {Spite}(1988)}]{cayrel1988}
{Cayrel de Strobel}, G., \& {Spite}, M., eds. 1988, IAU Symposium, Vol. 132,
  {The impact of very high S/N spectroscopy on stellar physics: proceedings of
  the 132nd Symposium of the International Astronomical Union held in Paris,
  France, June 29-July 3, 1987.}

\bibitem[{Coc {et~al.}(2004)Coc, Vangioni-Flam, Descouvemont, Adahchour, \&
  Angulo}]{coc2004}
Coc, A., Vangioni-Flam, E., Descouvemont, P., Adahchour, A., \& Angulo, C.
  2004, The Astrophysical Journal, 600, 544, \dodoi{10.1086/380121}

\bibitem[{Cohen {et~al.}(2004)Cohen, Christlieb, McWilliam, Shectman, Thompson,
  Wasserburg, Ivans, Dehn, Karlsson, \& Melendez}]{cohen2004}
Cohen, J.~G., Christlieb, N., McWilliam, A., {et~al.} 2004, The Astrophysical
  Journal, 612, 1107, \dodoi{10.1086/422576}

\bibitem[{D'Antona {et~al.}(2019)D'Antona, Ventura, Marino, Milone, Tailo,
  Criscienzo, \& Vesperini}]{dantona2019}
D'Antona, F., Ventura, P., Marino, A.~F., {et~al.} 2019, The Astrophysical
  Journal, 871, L19, \dodoi{10.3847/2041-8213/aafbec}

\bibitem[{{Dehnen} \& {Binney}(1998)}]{dehnenbinney98}
{Dehnen}, W., \& {Binney}, J. 1998, \mnras, 294, 429,
  \dodoi{10.1046/j.1365-8711.1998.01282.x}

\bibitem[{Denissenkov {et~al.}(2017)Denissenkov, Herwig, Battino, Ritter,
  Pignatari, Jones, \& Paxton}]{den2017}
Denissenkov, P.~A., Herwig, F., Battino, U., {et~al.} 2017, The Astrophysical
  Journal, 834, L10, \dodoi{10.3847/2041-8213/834/2/l10}

\bibitem[{Di~Matteo {et~al.}(2020)Di~Matteo, Spite, Haywood, Bonifacio, Gómez,
  Spite, \& Caffau}]{dimatteo2020}
Di~Matteo, P., Spite, M., Haywood, M., {et~al.} 2020, Astronomy \&
  Astrophysics, 636, A115, \dodoi{10.1051/0004-6361/201937016}

\bibitem[{D’Orazi {et~al.}(2015)D’Orazi, Gratton, Angelou, Bragaglia,
  Carretta, Lattanzio, Lucatello, Momany, Sollima, \& Beccari}]{dorazi2015}
D’Orazi, V., Gratton, R.~G., Angelou, G.~C., {et~al.} 2015, Monthly Notices
  of the Royal Astronomical Society, 449, 4038–4047,
  \dodoi{10.1093/mnras/stv612}

\bibitem[{{Eisenstein} {et~al.}(2011){Eisenstein}, {Weinberg}, {Agol},
  {Aihara}, {Allende Prieto}, {Anderson}, {Arns}, {Aubourg}, {Bailey},
  {Balbinot}, \& et~al.}]{eisenstein}
{Eisenstein}, D.~J., {Weinberg}, D.~H., {Agol}, E., {et~al.} 2011, \aj, 142,
  72, \dodoi{10.1088/0004-6256/142/3/72}

\bibitem[{{Frebel}(2018)}]{frebelrev18}
{Frebel}, A. 2018, Annual Review of Nuclear and Particle Science, 68, 237,
  \dodoi{10.1146/annurev-nucl-101917-021141}

\bibitem[{{Frebel} \& {Norris}(2015)}]{frebelandnorris}
{Frebel}, A., \& {Norris}, J.~E. 2015, \araa, 53, 631,
  \dodoi{10.1146/annurev-astro-082214-122423}

\bibitem[{Gallagher {et~al.}(2010)Gallagher, Ryan, García~Pérez, \&
  Aoki}]{gallagher2010}
Gallagher, A.~J., Ryan, S.~G., García~Pérez, A.~E., \& Aoki, W. 2010,
  Astronomy \& Astrophysics, 523, A24, \dodoi{10.1051/0004-6361/201014970}

\bibitem[{{Ge} {et~al.}(2015){Ge}, {Thomas}, {Li}, {Senan Seieroe Grieves},
  {Ma}, {de Lee}, {Lee}, {Liu}, {Bolton}, {Thakar}, {Weaver}, \& {SDSS-Iii
  Marvels Team}}]{ge2015}
{Ge}, J., {Thomas}, N.~B., {Li}, R., {et~al.} 2015, in American Astronomical
  Society Meeting Abstracts, Vol. 225, American Astronomical Society Meeting
  Abstracts \#225, 409.03

\bibitem[{{Hampel} {et~al.}(2016){Hampel}, {Stancliffe}, {Lugaro}, \&
  {Meyer}}]{hampel2016}
{Hampel}, M., {Stancliffe}, R.~J., {Lugaro}, M., \& {Meyer}, B.~S. 2016, \apj,
  831, 171, \dodoi{10.3847/0004-637X/831/2/171}

\bibitem[{Haywood {et~al.}(2018)Haywood, Matteo, Lehnert, Snaith, Khoperskov,
  \& G{\'{o}}mez}]{haywood2018}
Haywood, M., Matteo, P.~D., Lehnert, M.~D., {et~al.} 2018, The Astrophysical
  Journal, 863, 113, \dodoi{10.3847/1538-4357/aad235}

\bibitem[{{Heger} \& {Woosley}(2002)}]{hegerandwoosley2002}
{Heger}, A., \& {Woosley}, S.~E. 2002, \apj, 567, 532, \dodoi{10.1086/338487}

\bibitem[{{Helmi} {et~al.}(2018){Helmi}, {Babusiaux}, {Koppelman}, {Massari},
  {Veljanoski}, \& {Brown}}]{helmi2018}
{Helmi}, A., {Babusiaux}, C., {Koppelman}, H.~H., {et~al.} 2018, \nat, 563, 85,
  \dodoi{10.1038/s41586-018-0625-x}

\bibitem[{{Korn} {et~al.}(2006){Korn}, {Grundahl}, {Richard}, {Barklem},
  {Mashonkina}, {Collet}, {Piskunov}, \& {Gustafsson}}]{korn2006}
{Korn}, A.~J., {Grundahl}, F., {Richard}, O., {et~al.} 2006, \nat, 442, 657,
  \dodoi{10.1038/nature05011}

\bibitem[{{Lai} {et~al.}(2008){Lai}, {Bolte}, {Johnson}, {Lucatello}, {Heger},
  \& {Woosley}}]{lai2008}
{Lai}, D.~K., {Bolte}, M., {Johnson}, J.~A., {et~al.} 2008, \apj, 681, 1524,
  \dodoi{10.1086/588811}

\bibitem[{{Matas Pinto} {et~al.}(2021){Matas Pinto}, {Spite}, {Caffau},
  {Bonifacio}, {Sbordone}, {Sivarani}, {Steffen}, {Spite}, {Fran{\c{c}}ois}, \&
  {Di Matteo}}]{pinto2021}
{Matas Pinto}, A.~M., {Spite}, M., {Caffau}, E., {et~al.} 2021, \aap, 654,
  A170, \dodoi{10.1051/0004-6361/202141288}

\bibitem[{{McMillan}(2017)}]{mcmillan2017}
{McMillan}, P.~J. 2017, \mnras, 465, 76, \dodoi{10.1093/mnras/stw2759}

\bibitem[{{McWilliam}(1998)}]{mcwilliam1998}
{McWilliam}, A. 1998, \aj, 115, 1640, \dodoi{10.1086/300289}

\bibitem[{{Meingast} {et~al.}(2021){Meingast}, {Alves, Jo\~ao}, \&
  {Rottensteiner, Alena}}]{meingast21}
{Meingast}, {Alves, Jo\~ao}, \& {Rottensteiner, Alena}. 2021, A\&A, 645, A84,
  \dodoi{10.1051/0004-6361/202038610}

\bibitem[{{Nakamura} {et~al.}(1999){Nakamura}, {Umeda}, {Nomoto}, {Thielemann},
  \& {Burrows}}]{nakamura1999}
{Nakamura}, T., {Umeda}, H., {Nomoto}, K., {Thielemann}, F.-K., \& {Burrows},
  A. 1999, \apj, 517, 193, \dodoi{10.1086/307167}

\bibitem[{{Nomoto} {et~al.}(2013){Nomoto}, {Kobayashi}, \&
  {Tominaga}}]{nomoto2013}
{Nomoto}, K., {Kobayashi}, C., \& {Tominaga}, N. 2013, \araa, 51, 457,
  \dodoi{10.1146/annurev-astro-082812-140956}

\bibitem[{{Pasquini} {et~al.}(2005){Pasquini}, {Bonifacio}, {Molaro},
  {Francois}, {Spite}, {Gratton}, {Carretta}, \& {Wolff}}]{pasquini}
{Pasquini}, L., {Bonifacio}, P., {Molaro}, P., {et~al.} 2005, \aap, 441, 549,
  \dodoi{10.1051/0004-6361:20053607}

\bibitem[{{Piau} {et~al.}(2006){Piau}, {Beers}, {Balsara}, {Sivarani},
  {Truran}, \& {Ferguson}}]{piau2006}
{Piau}, L., {Beers}, T.~C., {Balsara}, D.~S., {et~al.} 2006, \apj, 653, 300,
  \dodoi{10.1086/508445}

\bibitem[{{Pinsonneault} {et~al.}(1999){Pinsonneault}, {Walker}, {Steigman}, \&
  {Narayanan}}]{Pinsonneault}
{Pinsonneault}, M.~H., {Walker}, T.~P., {Steigman}, G., \& {Narayanan}, V.~K.
  1999, \apj, 527, 180, \dodoi{10.1086/308048}

\bibitem[{{Placco} {et~al.}(2014){Placco}, {Frebel}, {Beers}, {Christlieb},
  {Lee}, {Kennedy}, {Rossi}, \& {Santucci}}]{placco2014}
{Placco}, V.~M., {Frebel}, A., {Beers}, T.~C., {et~al.} 2014, \apj, 781, 40,
  \dodoi{10.1088/0004-637X/781/1/40}

\bibitem[{{Roederer} {et~al.}(2014){Roederer}, {Preston}, {Thompson},
  {Shectman}, \& {Sneden}}]{roederer2014}
{Roederer}, I.~U., {Preston}, G.~W., {Thompson}, I.~B., {Shectman}, S.~A., \&
  {Sneden}, C. 2014, \apj, 784, 158, \dodoi{10.1088/0004-637X/784/2/158}

\bibitem[{Ruchti {et~al.}(2011)Ruchti, Fulbright, Wyse, Gilmore, Grebel,
  Bienaym{\'{e}}, Bland-Hawthorn, Freeman, Gibson, Munari, Navarro, Parker,
  Reid, Seabroke, Siebert, Siviero, Steinmetz, Watson, Williams, \&
  Zwitter}]{ruchti2011}
Ruchti, G.~R., Fulbright, J.~P., Wyse, R. F.~G., {et~al.} 2011, The
  Astrophysical Journal, 743, 107, \dodoi{10.1088/0004-637x/743/2/107}

\bibitem[{Ryan {et~al.}(2002)Ryan, Gregory, Kolb, Beers, \& Kajino}]{ryan2002}
Ryan, S.~G., Gregory, S.~G., Kolb, U., Beers, T.~C., \& Kajino, T. 2002, The
  Astrophysical Journal, 571, 501, \dodoi{10.1086/339939}

\bibitem[{{Ryan} {et~al.}(1999){Ryan}, {Norris}, \& {Beers}}]{ryan1999}
{Ryan}, S.~G., {Norris}, J.~E., \& {Beers}, T.~C. 1999, \apj, 523, 654,
  \dodoi{10.1086/307769}

\bibitem[{Sbordone {et~al.}(2010)Sbordone, Bonifacio, Caffau, Ludwig, Behara,
  González~Hernández, Steffen, Cayrel, Freytag, Van’t~Veer, \&
  et~al.}]{sbordone2010}
Sbordone, L., Bonifacio, P., Caffau, E., {et~al.} 2010, Astronomy \&
  Astrophysics, 522, A26, \dodoi{10.1051/0004-6361/200913282}

\bibitem[{Siegel {et~al.}(2019)Siegel, Barnes, \& Metzger}]{siegel2019}
Siegel, D.~M., Barnes, J., \& Metzger, B.~D. 2019, Nature, 569, 241–244,
  \dodoi{10.1038/s41586-019-1136-0}

\bibitem[{Spergel {et~al.}(2003)Spergel, Verde, Peiris, Komatsu, Nolta,
  Bennett, Halpern, Hinshaw, Jarosik, Kogut, Limon, Meyer, Page, Tucker,
  Weiland, Wollack, \& Wright}]{spergel2003}
Spergel, D.~N., Verde, L., Peiris, H.~V., {et~al.} 2003, The Astrophysical
  Journal Supplement Series, 148, 175, \dodoi{10.1086/377226}

\bibitem[{{Spite} \& {Spite}(1982)}]{spitenspite}
{Spite}, F., \& {Spite}, M. 1982, \aap, 115, 357

\bibitem[{{Spite} {et~al.}(2005){Spite}, {Cayrel}, {Plez}, {Hill}, {Spite},
  {Depagne}, {Fran{\c c}ois}, {Bonifacio}, {Barbuy}, {Beers}, {Andersen},
  {Molaro}, {Nordstr{\"o}m}, \& {Primas}}]{firststars6}
{Spite}, M., {Cayrel}, R., {Plez}, B., {et~al.} 2005, \aap, 430, 655,
  \dodoi{10.1051/0004-6361:20041274}

\bibitem[{{Spite, M.} {et~al.}(2015){Spite, M.}, {Spite, F.}, {Caffau, E.}, \&
  {Bonifacio, P.}}]{litospite}
{Spite, M.}, {Spite, F.}, {Caffau, E.}, \& {Bonifacio, P.} 2015, A\&A, 582,
  A74, \dodoi{10.1051/0004-6361/201526878}

\bibitem[{Sriram {et~al.}(2018)Sriram, Kumar, Surya, Sivarani, Giridhar,
  Kathiravan, Anand, Jones, Grobler, Jakobsson, Chanumolu, Unni, Dorje, Dorje,
  \& Gyalson}]{sriram2018}
Sriram, S., Kumar, A., Surya, A., {et~al.} 2018, in Ground-based and Airborne
  Instrumentation for Astronomy VII, ed. C.~J. Evans, L.~Simard, \& H.~Takami,
  Vol. 10702, International Society for Optics and Photonics (SPIE), 2007 --
  2021.
\newblock \url{https://doi.org/10.1117/12.2313165}

\bibitem[{{Suda} {et~al.}(2008){Suda}, {Katsuta}, {Yamada}, {Suwa}, {Ishizuka},
  {Komiya}, {Sorai}, {Aikawa}, \& {Fujimoto}}]{sudasaga}
{Suda}, T., {Katsuta}, Y., {Yamada}, S., {et~al.} 2008, \pasj, 60, 1159,
  \dodoi{10.1093/pasj/60.5.1159}

\bibitem[{Susmitha {et~al.}(2021)Susmitha, Ojha, Sivarani, Ninan,
  Bandyopadhyay, Surya, \& Unni}]{susmitha2021}
Susmitha, A., Ojha, D.~K., Sivarani, T., {et~al.} 2021, Monthly Notices of the
  Royal Astronomical Society, 506, 1962, \dodoi{10.1093/mnras/stab1508}

\bibitem[{{Susmitha Rani} {et~al.}(2016){Susmitha Rani}, {Sivarani}, {Beers},
  {Fleming}, {Mahadevan}, \& {Ge}}]{susmitha}
{Susmitha Rani}, A., {Sivarani}, T., {Beers}, T.~C., {et~al.} 2016, \mnras,
  458, 2648, \dodoi{10.1093/mnras/stw413}

\bibitem[{{Truran} \& {Arnett}(1971)}]{truran1971}
{Truran}, J.~W., \& {Arnett}, W.~D. 1971, \apss, 11, 430,
  \dodoi{10.1007/BF00649636}

\bibitem[{{Tsujimoto} \& {Shigeyama}(2014{\natexlab{a}})}]{tsuji1}
{Tsujimoto}, T., \& {Shigeyama}, T. 2014{\natexlab{a}}, \apjl, 795, L18,
  \dodoi{10.1088/2041-8205/795/1/L18}

\bibitem[{{Tsujimoto} \& {Shigeyama}(2014{\natexlab{b}})}]{tsuji2}
---. 2014{\natexlab{b}}, \aap, 565, L5, \dodoi{10.1051/0004-6361/201423751}

\bibitem[{{Umeda} {et~al.}(2000){Umeda}, {Nomoto}, \& {Nakamura}}]{umeda2000}
{Umeda}, H., {Nomoto}, K., \& {Nakamura}, T. 2000, in The First Stars, ed.
  A.~{Weiss}, T.~G. {Abel}, \& V.~{Hill}, 150, \dodoi{10.1007/10719504_27}

\end{thebibliography}


\end{document}